\documentclass[fleqn,usenatbib]{mnras}

\usepackage[T1]{fontenc}

\DeclareRobustCommand{\VAN}[3]{#2}
\let\VANthebibliography\thebibliography
\def\thebibliography{\DeclareRobustCommand{\VAN}[3]{##3}\VANthebibliography}

\usepackage{graphicx}	
\usepackage{amsmath}	
\usepackage{amssymb}	
\usepackage{physics}
\usepackage{comment}

\newcommand\T{\rule{0pt}{2.6ex}}       
\newcommand\B{\rule[-1.2ex]{0pt}{0pt}} 

\newcommand{\swift}{\textsc{Swift}}
\newcommand{\sphenix}{\textsc{sphenix}}

\usepackage{newtxtext,newtxmath}

\title[FLAMINGO: Calibration with machine learning]{FLAMINGO: Calibrating large cosmological hydrodynamical simulations with machine learning}

\author[R. Kugel et al.]{
Roi Kugel,$^{1}$\thanks{E-mail: kugel@strw.leidenuniv.nl}
Joop Schaye,$^{1}$
Matthieu Schaller,$^{2,1}$
John C. Helly,$^{3}$
Joey Braspenning,$^{1}$
Willem Elbers,$^{3}$\newauthor
Carlos S. Frenk,$^{3}$
Ian G. McCarthy,$^{4}$
Juliana Kwan,$^{4}$
Jaime Salcido,$^{4}$
Marcel P. van Daalen,$^{1}$\newauthor
Bert Vandenbroucke,$^{1}$
Yannick M. Bahé,$^{1,5}$
Josh Borrow,$^{3,6}$
Evgenii Chaikin,$^{1}$
Filip Hu\v{s}ko,$^{3}$\newauthor
Adrian Jenkins,$^{3}$
Cedric G. Lacey,$^{3}$
Folkert S. J. Nobels,$^{1}$
and Ian Vernon$^{7}$
\\
$^{1}$Leiden Observatory, Leiden University, PO Box 9513, 2300 RA Leiden, the Netherlands\\
$^{2}$Lorentz Institute for Theoretical Physics, Leiden University, PO box 9506, 2300 RA Leiden, the Netherlands\\
$^{3}$Institute for Computational Cosmology, Department of Physics, University of Durham, South Road, Durham, DH1 3LE, UK\\
$^{4}$Astrophysics Research Institute, Liverpool John Moores University, Liverpool L3 5RF, UK\\
$^{5}$Institute of Physics, Laboratory of Astrophysics, Ecole Polytechnique Fédérale de Lausanne (EPFL), Observatoire de Sauverny, 1290 Versoix, Switzerland\\
$^{6}$Department of Physics, Kavli Institute for Astrophysics and Space Research, Massachusetts Institute of Technology, Cambridge, MA 02139, USA\\
$^{7}$Department of Mathematical Sciences, Durham University,
Stockton Road, DH1 3LE, Durham, UK
}

\date{Accepted XXX. Received YYY; in original form ZZZ}

\pubyear{2022}

\begin{document}
\label{firstpage}
\pagerange{\pageref{firstpage}--\pageref{lastpage}}
\maketitle

\begin{abstract}
To fully take advantage of the data provided by large-scale structure surveys, we need to quantify the potential impact of baryonic effects, such as feedback from active galactic nuclei (AGN) and star formation, on cosmological observables. In simulations, feedback processes originate on scales that remain unresolved. Therefore, they need to be sourced via subgrid models that contain free parameters. We use machine learning to calibrate the AGN and stellar feedback models for the FLAMINGO cosmological hydrodynamical simulations. Using Gaussian process emulators trained on Latin hypercubes of 32 smaller-volume simulations, we model how the galaxy stellar mass function and cluster gas fractions change as a function of the subgrid parameters. The emulators are then fit to observational data, allowing for the inclusion of potential observational biases. We apply our method to the three different FLAMINGO resolutions, spanning a factor of 64 in particle mass, recovering the observed relations within the respective resolved mass ranges. We also use the emulators, which link changes in subgrid parameters to changes in observables, to find models that skirt or exceed the observationally allowed range for cluster gas fractions and the stellar mass function. Our method enables us to define model variations in terms of the data that they are calibrated to rather than the values of specific subgrid parameters. This approach is useful, because subgrid parameters are typically not directly linked to particular observables, and predictions for a specific observable are influenced by multiple subgrid parameters.
\end{abstract}




\section{Introduction}
The evolution of the large-scale distribution of matter in the Universe is highly sensitive to the underlying cosmological model. Current probes have given us our concordance cosmological model $\Lambda$CDM, which consists of a spatially flat universe, where dark energy and cold dark matter dominate the current energy density \citep[for a review see][]{DEReview2008}.

The concordance model has been independently validated by a large array of probes. These include the cosmic microwave background (CMB)  \citep[e.g.][]{Planck2020}, galaxy clustering and gravitational lensing \citep[e.g.][]{DESYR3,KIDS2021}, baryon acoustic oscillations (BAO)  \citep[e.g.][]{BOSS2021}, and more \citep[for a review see][]{Turner2022}. While all the probes broadly agree with the $\Lambda$CDM model, tensions remain between early universe probes, like the CMB, and late-time probes, like the distance ladder and weak lensing. For the $H_0$ and $\sigma_8$ parameters, the tension is at the level of a few standard deviations \citep[e.g.][]{KIDS2021,DESYR3,Ries2022}. Next generation surveys like \textit{Euclid}\footnote{\url{https://www.euclid-ec.org/}} and LSST\footnote{\url{https://www.lsst.org/}} will measure the matter power spectrum to per cent level accuracy \citep{Euclidprep2020}. The results from these surveys will provide us with a stringent test of the concordance model, and show us whether these tensions will force us to modify the $\Lambda$CDM model.

Most of the modelling work for large-scale structure is done with collisionless $N$-body simulations \citep[e.g.][]{MiraTitan,DESyr3Cosmosims,EUCLIDemu}. $N$-body simulations model the evolution of cold dark matter and can accurately predict the structure and clustering of dark matter haloes under the effect of gravity only. The dark part of the matter component is dominant in mass and hence, predictions from these simulations may provide stringent cosmological constraints. However, baryons change the distribution of dark matter through back reaction effects, but, with the exception of gravitational lensing, we are limited to observing the imprint of the distribution of dark matter on the baryonic matter. Most of the baryonic matter is found in the tenuous intergalactic medium \citep[e.g.][]{IGM2018,IGMNatureMacquart2020}, which is very challenging to observe directly. Large-scale structure surveys use galaxies, which are located within dark matter haloes, to map the distribution of matter. 

Sophisticated semi-analytical and semi-empirical models can make predictions for how galaxies evolve within their dark matter haloes \citep[e.g.][]{Galform2015,GALFORM16,Moster2018,Behroozi2019,Lgalaxies2021}. Baryonic effects can be simulated with halo models \citep[e.g.][]{Semboloni2011,Semboloni2013,Mead2015,Stijn2020,Acuto2021}, added to N-body simulations by baryonification algorithms \citep[e.g.][]{Baryonification2015,Baryonification2_2021,Baryonifsim2021} or included as a parametric correction to the matter power spectrum \citep{vanDaalen2020,Salcido2023}. However, the most self-consistent way to model how the large-scale structure is coupled with baryons, is via large cosmological hydrodynamical simulations. Modern simulations like Magneticum \citep{Magneticum2014}, EAGLE \citep{Eaglemain,EagleCal}, Horizon-AGN \citep{HorizonAGN2017}, IllustrisTNG \citep{Illustris2018}, BAHAMAS \citep{BAHAMAS2017,BAHAMAS2018}, SIMBA \cite{SIMBA2019} and MilleniumTNG \citep{MTNG} provide predictions for the interplay between galaxy formation and the large-scale structure. The results from hydrodynamical simulations can also inform the simpler parametric and analytic models.

One of the main difficulties for hydrodynamical simulations is the implementation and tuning of relevant astrophysical processes that originate on unresolved scales through subgrid physics. Processes like star formation and black hole growth occur on parsec scales, and are not resolved. The resulting feedback from stars and active galactic nuclei (AGN), do influence the distribution of matter on cosmological scales \citep{VanDaalen2011,vanDaalen2020,Stijn2020,Schneider2020}. Therefore, we need to create simulations that model their effect on the resolved scales. 

Subgrid physics models are characterised by a set of free parameters, in the sense that there is both uncertainty in the processes we try to model and uncertainty in how the models are affected by numerical limitations. An example of the latter is the impact of numerical over-cooling on galactic wind models \citep[see][]{DVSchaye2012Therm}. The numerical effects combined with the general non-linearity of galaxy formation makes it difficult to implement subgrid physics based solely on first principles. Instead, we have to calibrate the model by comparing it to a selection of observations, a partial forfeit of their predictive power. As argued by \cite{Eaglemain}, this is a necessary sacrifice. By ensuring certain relations are reproduced, the simulation retains predictive power for other relations. Calibrating subgrid physics forces us to find a balance between how many observables one tries to match and how many of the results can be deemed predictions.

In this paper we discuss the calibration strategy used for the low-, intermediate- and high-resolution simulations of the FLAMINGO project (Full-hydro Large-scale structure simulations with All-sky Mapping for the 
Interpretation of Next Generation Observations; \citealt{Flamingomain}). The intermediate-resolution FLAMINGO model has the same resolution ($m_{\rm gas}=1.07\times10^{9}~\rm{M}_{\odot}$) as used for the BAHAMAS project \citep{BAHAMAS2017,BAHAMAS2018}, but in a volume of $(2.8~\rm{Gpc})^3$. This volume is over two orders of magnitude larger than BAHAMAS. Additionally, FLAMINGO includes a suite of feedback and cosmology variations in $(1~\rm{Gpc})^3$ volumes. This includes a high ($m_{\rm gas}=1.34\times10^{8}~\rm{M}_{\odot}$) and a low ($m_{\rm gas}=8.56\times10^{9}~\rm{M}_{\odot}$) resolution variation. Our goal is to expand the large-scale structure science of the BAHAMAS project to larger volumes, different resolutions, and more cosmology and astrophysics variations with a new code and an improved subgrid physics model. The FLAMINGO simulation outputs also include on-the-fly full sky lightcones, both as particles and as maps, for a variety of observables. Similarly to BAHAMAS, we will calibrate to the observed present-day galaxy stellar mass function (SMF) and the gas fractions in groups and clusters of galaxies ($f_\text{gas}$). We opt for the SMF to ensure we can reproduce galaxy clustering and lensing statistics if we use the correct cosmology. The gas fraction is used to ensure we have a realistic distribution of gas in and around clusters, which is not only important for cluster cosmology, but also for baryonic effects on the matter power spectrum \citep{Semboloni2011,Baryonification2015,Stijn2020,vanDaalen2020,Baryonifsim2021,Salcido2023}. While our fiducial models are calibrated to reproduce the data, we also calibrate the subgrid physics to the gas fraction and SMF data that has been shifted relative to the observed values. These feedback variations will enable future FLAMINGO projects to test the importance of astrophysical effects constrained by the uncertainties in the data.

For BAHAMAS, and also for simulations like EAGLE and IllustrisTNG, calibration was done by hand by varying the subgrid parameters within some reasonable range until the simulation lined up with the calibration targets. This approach works reasonably well in the context of galaxy formation, but it introduces biases into the parameter selection. For cosmology applications we require a more systematic and controlled approach. We want to be able to sample the parameter space with a Markov Chain Monte Carlo (MCMC) method and to find the posterior probabilities of each of the subgrid parameter values. This approach also allows us to take into account potential systematic effects in the data and/or simulations.

Because N-body simulations are too computationally expensive to be used directly in MCMC-like methods, we make use of machine learning, specifically emulation using Gaussian processes. While it is too expensive to run a new simulation for each MCMC step, we can train an emulator on a carefully sampled selection of input simulations. The emulator then gives us the predicted observable as a continuous function of the input parameters, which can be fed into any likelihood calculation code. Emulator-based methods have been used in combination with semi-analytic models of galaxy formation \citep{Bower2010,Vernon2014,Rodrigues2017,Galformemu21} and have become particularly popular for cosmology. By training emulators on dark-matter-only simulations, their full non-linear matter power spectrum can be predicted with per cent level precision \citep[e.g.][]{Heitmann2009a,Heitmann2016,EUCLIDemu,Angulo2021,Moran2022}.

We directly emulate our calibration targets: the SMF and the gas fractions in groups and clusters. This allows us to create a continuous simulation-based model that can be compared with observations. With the emulator we can use MCMC to directly fit the subgrid physics parameters to the observational data, while modelling statistical and systematic errors in both the simulations and the data. This procedure not only gives us a well-calibrated model, but also lets us determine the maximum variations allowed by the model. In this way our resulting simulations can provide upper and lower limits on the expected baryonic effects. More general machine learning techniques have been used to calibrate hydrodynamical simulations. \cite{HydroMLCal2} calibrate to baryonic observables in the $(25~\rm{Mpc})^3$ volumes of the CAMELS project \citep{CAMELS} and \cite{HydroMLCal1} apply a similar methodology to zooms of Milky Way haloes. However, these methods have not been applied to simulations of large cosmological volumes and they have not accounted for possible observational biases.

This paper is structured as follows. In Section~\ref{sec:simulations} we describe the most relevant aspects of our simulation method and galaxy formation models. In Section~\ref{sec:obsdata} the reasoning for our calibration targets is explained, and we describe our compilation of data and how we include potential observational and simulation-originated biases in our analysis.  In Section~\ref{sec:emulator_descr} we describe how we obtain the training data for the emulators. We also discuss how the emulators are trained and how we estimate the uncertainty in the predictions of the emulators. We describe our likelihoods and our fitting method in Section~\ref{sec:likelihoods}. In Section~\ref{sec:Results} we show the results of fitting the emulators at the three FLAMINGO resolutions. We also discuss how the emulators can be used to better understand subgrid physics using parameter sweeps and we use the emulator to find models that skirt or exceed the observational allowed range for the cluster gas fractions and the SMF. Finally, we summarise our method, strategy and results in Section~\ref{sec:conc}. In this work, $R_{500c}$ is defined as the radius within which the mean internal density is 500 times the critical density. The radius $R_{500c}$ also defines $M_{500c}$, which is the mass inside $R_{500c}$. 

\section{Simulations} \label{sec:simulations}
The simulation methods and galaxy formation model are described in detail in \citet{Flamingomain}. Here we will provide a summary of the most relevant aspects. We describe in more detail the subgrid prescriptions that we calibrate in this work, namely those for stellar feedback (\S\ref{sec:stellar_feedback}), the growth of supermassive black holes (\S\ref{sec:black_holes}), and AGN feedback (\S\ref{Sec:agn_feedback}), and we will motivate the choice of priors for the subgrid parameters that are varied (these are listed in Table~\ref{tab:paramtab}). 

All simulations in this work use the open-source code \swift\ \citep{SWIFT}. \swift\ is an N-body gravity and smooth particle hydrodynamics (SPH) solver that makes use of a fine-grained tasking framework and runs across multiple compute nodes using \textsc{MPI}. Gravity is solved using the Fast Multiple Method \citep{Greengard1987AFA}. We use the \sphenix\ SPH scheme \citep{Borrow2022Sphenix} with a \cite{wendland1995} $C^2$ kernel. Massive neutrinos are implemented into \swift\ via the $\delta f$
method of \cite{Elbers2021}. 

Initial conditions are generated using a modified version of \textsc{monofonIC} \citep{Han2021} that includes massive neutrinos. We use unperturbed initial conditions for the neutrino particles. We do not include large scale neutrino perturbations in the initial conditions, as these have a negligible effect in the small box sizes used for this work. We adopt the '3x2pt + all' cosmology from \cite{DESYR3} $(\Omega_{\rm m} = 0.306, \ \Omega_{\rm b} = 0.0486, \ \sigma_8 = 0.807, \ {\rm H}_{0} = 68.1, \ n_{s} = 0.967)$ with a minimal neutrino mass of 0.06 eV. The particle masses and gravitational softening lengths corresponding to the three different resolutions that we will consider are listed in Table~\ref{tab:resolutions}.

\begin{table*}
	\centering
	\caption{Numerical characteristics of the final Latin hypercubes of simulations. The columns list: the resolution qualifier, comoving box size, number of particles (there are initially equal numbers of dark matter and baryonic particles), initial baryonic particle mass, dark matter particle mass, comoving gravitational softening length, maximum physical gravitational softening length.}
	\label{tab:resolutions}
	\begin{tabular}{lrccccr}
		\hline
		Resolution & $L$ & $N$ & $m_\text{g}$ & $m_\text{DM}$ & $\epsilon_\text{com}$ & $\epsilon_\text{prop}$\\
		           & (cMpc) && (M$_\odot$) & (M$_\odot$)  & (ckpc) & (pkpc) \\
		\hline
		Low [m10]         & 400 & $2\times 360^3$ & $8.56\times 10^9$ & $4.52\times 10^{10}$ & $44.6$ & $11.40$ \\
		Intermediate [m9] & 200 & $2\times 360^3$ & $1.07\times 10^9$ & $5.65\times 10^9$ & 22.3 & 5.70\\
		High [m8]        & 100& $2\times 360^3$ & $1.34\times 10^8$ & $7.06\times 10^8$ & 11.2 & 2.85\\
		\hline
	\end{tabular}
\end{table*}

For simulations with volumes as large as FLAMINGO, it is currently impossible to resolve all the processes that are important for galaxy formation. Therefore, we make use of subgrid models. FLAMINGO builds upon the models of OWLS \citep{owls}, used for Cosmo-OWLS \citep{OWLS2014}, BAHAMAS \citep{BAHAMAS2017}, and EAGLE \citep{Eaglemain}, ported from the code \textsc{gadget} \citep{Springel2005} to \swift.

We use the radiative cooling tables from \cite{Ploeckinger2020}, which are based on photo-ionisation models run with \textsc{cloudy} \citep{Ferland2017} that include both the metagalactic and interstellar radiation fields, and that account for self-shielding, dust, and cosmic rays. 

As we are unable to resolve the multiphase interstellar medium, we follow \citet{SchayeDV2008} and impose a temperature floor. The pressure of gas with hydrogen number densities $n_{\rm H}>10^{-4}  \ {\rm cm}^{-3}$ and an overdensity greater than 100 is limited from below to $P/k_{\rm B} = 800 \ {\rm K}  \ (n_{\rm H} / 10^{-4} \ {\rm cm}^{-3})^{4/3}$, where $k_{\rm B}$ is the Boltzmann constant.

During the simulation gas particles can be stochastically converted into star particles following the description of \cite{SchayeDV2008}. Particles with total hydrogen number density\footnote{Due to a bug, in the intermediate-resolution simulations gas particles with a metallicity equal to exactly zero were only allowed to form stars at densities higher than $10~{\rm cm}^{-3}$. This had little to no effect on any results at resolved stellar masses, but it did reduce the number of stars formed in the lowest-mass galaxies. Fixing this bug would potentially have allowed us to match the SMF to stellar masses corresponding to fewer than 10 particles. \label{footnote:sf_bug}} $n_{\rm H}>10^{-1} \  {\rm cm}^{-3}$, an overdensity $>10$ and within 0.3 dex of the temperature floor are stochastically allowed to convert into stars with a probability given by the particle's star formation rate, 
\begin{equation}
    \dot m_{*} = m_\text{g} A (1 ~{\rm M}_{\odot} {\rm pc}^{-2} )^{-n} \left(\frac{\gamma}{G}f_\text{g} P\right)^{(n-1)/2},
\end{equation} 
where $m_\text{g}$ is the gas particle mass, $\gamma = 5/3$ is the adiabatic index, and $G$ is the gravitational constant. The star formation rate is derived such that self-gravitating discs reproduce the observed Kennicutt-Schmidt relation \citep{Kennicutt1998,Kennicutt2007}. We assume the gas fraction, $f_\text{g}$, is unity, $A = 1.515\times 10^{-4} \ {\rm M}_{\odot}~{\rm yr}^{-1}~{\rm pc}^{-2}$, and $n=1.4$.

For the low-resolution simulation we were forced to relax the star formation parameters, as the default prescription was unable to form enough stars, even in large haloes and without stellar feedback. For low resolution, all particles with density $n_{\rm H}>10^{-3} \ {\rm cm}^{-3}$, overdensity $>10$ and temperature $T<10^5 \  {\rm K}$ are star forming. 

Each stellar particle is treated as a simple stellar population with  a \citet{Chabrier2003} initial mass function (IMF). Following \citet{Wiersma2009chemistry}, we model stellar mass loss and track the abundances of the individual elements H, He, C, N, O, Ne, Mg, Si, and Fe. We also include type Ia supernova with rates taken from \cite{Eaglemain}.

\begin{table*}
    \centering
    \caption{Priors and best-fitting values for the subgrid parameters for each of the three simulation resolutions. Low-resolution simulations do not include stellar feedback. The rows titled 'Median+CL' give the median and the 16th and 84th percentile confidence level (CL) obtained from the posterior of the fits. The rows titled 'best-fitting' list the maximum likelihood value from the fitting, which is our fiducial value. The last row 'Log' indicates whether the parameter is sampled logarithmically. The best-fitting values for the jet model are listed in Table~\ref{tab:variations} and the priors for the jet model are listed in Table~\ref{tab:jet_params}.}
    \begin{tabular}{l|l|l|l|l|l}
    \hline Resolution & Parameter & $f_{\rm SN}$ & $\Delta v_\text{SN}$ & $\log_{10}\Delta T_\text{AGN}$ [K] & $\beta_{\text{BH}}$ \\
    \hline & Prior & $[0.2,0.9]$ & $[80,400]$ & $[7.7,8.9]$ & $[0.0,0.9]$ \\
    High-res [m8] & Median+CL & $0.56^{0.15}_{-0.12}$ & $169^{+87}_{-65}$ & $8.03^{+0.13}_{-0.14}$ & $0.23^{0.20}_{-0.15}$ \\
    & best-fitting & $0.524$ & $259$ & $8.07$ & $0.038$\\
    \hline
    & Prior & $[0,0.5]$ & $[200,800]$ & $[7.5,8.5]$ & $[0.1,0.9]$ \\
    Intermediate-res [m9] & Median+CL & $0.20^{+0.11}_{-0.09}$ & $479^{+167}_{-197}$ & $7.84^{+0.18}_{0.20}$ & $0.55^{+0.15}_{-0.16}$ \\
    & best-fitting & $0.238$ & $562$ & $7.95$ & $0.514$ \\
    \hline
    & Prior & - & - & $[7,9.5]$ & $[0,3]$ \\
    Low-res [m10] & Median+CL & - & - & $8.26^{+0.15}_{-0.15}$ & $0.50^{+0.17}_{-0.16}$ \\
    & best-fitting & - & - & $8.29$ & $0.373$ \\
    \hline
    & Log & No & Yes & Yes & No \\ \hline
    \end{tabular}
    \label{tab:paramtab}
\end{table*}

\subsection{Stellar feedback}\label{sec:stellar_feedback}
Although we will often refer to stellar feedback as supernova feedback, it may also represent other sources of energy released by massive stars that are unresolved by our simulations such as stellar winds, radiation pressure or cosmic rays.

Stellar feedback is implemented kinetically. The energy budget is normalised to the expected kinetic energy from core collapse supernovae, assuming that each star with a mass between 8 and 100 ${\rm M}_{\odot}$ injects $10^{51}$ erg of kinetic energy into its surrounding medium. A fraction $f_\text{SN}$ of this energy is assumed to be coupled to the ISM on scales resolved by the simulation and is used to kick neighbouring gas particles with a target velocity $\Delta v_\text{SN}$. We use the method of \citet{Chaikin2022b}\footnote{There is one difference w.r.t. the method described by the authors. In the case where a particle would be kicked twice in a single time step, which we do not allow, we put the unused kick energy in a thermal dump, instead of adding it back to the star's feedback energy reservoir.} to inject the kinetic energy in a statistically isotropic manner while ensuring that both momentum and energy are conserved. Note that if the relative velocities between the star and gas particles are nonzero, energy conservation results in differences between the actual and target kick velocities. 

Following \citet{DVSchaye2008kin} and \citet{Richings2016}, we inject the kinetic energy probabilistically during each time step after the star particle has formed. The probability that a star particle kicks a given SPH neighbour is
\begin{equation}
    p_{\rm kick} (f_{\rm SN},\Delta v_{\rm SN}, m_{\rm ngb}, t, \Delta t) = 2\frac{f_{\rm SN} \Delta E_{\rm SNII}(t,\Delta t)}{m_{\rm ngb} \Delta v_{\rm SN}^2},
\end{equation}
where $\Delta E_{\rm SN}$ denotes the amount of energy released by the star particle of age $t$ during a time step $\Delta t$ and $m_{\rm ngb}$ is the total gas mass in the star particle's SPH kernel. The feedback efficiency, $f_{\rm SN}$, and the target kick velocity $\Delta v_{\rm SN}$ are the two stellar feedback parameters that are varied during the calibration.

The effect of stellar feedback generally scales with $f_\text{SN}$, which sets the amount of energy that is injected. Based on the calibration of BAHAMAS \citep{BAHAMAS2017} and after some experimentation with runs in which we varied only one parameter, we settled on prior ranges of $0.2-0.9$ and $0-0.5$ for high- and intermediate-resolution, respectively. The low-resolution simulations do not require any stellar feedback at all because of the strong suppression of star formation due to the limited resolution and because galaxies in the regime where stellar feedback dominates (stellar mass $M_* \ll 10^{11}\,\text{M}_\odot$) are only sampled by $\lesssim 10$ stellar particles. 

If the kick velocity is too small, then stellar feedback ceases to be effective because of excessive radiative losses caused by the too-low post-shock temperatures \citep[the well-known numerical over-cooling problem, see][]{DVSchaye2012Therm} and/or because the velocities are small compared to the escape velocities. The lower limits for $\Delta v_\text{SN}$ are 80 and 200 km~s$^{-1}$ for the high- and intermediate-resolution simulations, respectively. Our additional tests showed that for lower velocities the kicks stopped having a significant effect.

If the kick velocity is too large, then the feedback becomes poorly sampled, thus limiting its effectiveness. Our aim is to calibrate the SMF down to masses corresponding to just a few stellar particles. The expectation value for the number of kicks imparted by a single stellar particle is given by \citet{Chaikin2022b}
\begin{equation}
    \langle N_{\rm kicks, \ SN} \rangle = 1.85 \left( \frac{f_{\rm SN}}{0.25}\right) \left(\frac{\Delta v_{\rm SN}}{400 \ {\rm km \ s}^{-1}}\right)^{-2},
\end{equation}
where we assumed the stellar and gas particles to have the same mass. Based on the above considerations and some small test runs, we limit the maximum kick velocity to 400 and 800 km~s$^{-1}$ for the high- and intermediate-resolution simulations, respectively. This implies $\langle N_{\rm kicks, \ SN} \rangle\approx2$ and $\langle N_{\rm kicks, \ SN} \rangle\approx0.4$ for high- and intermediate-resolution respectively. There should be at least four kicks for objects with 10 stellar particles at each resolution.

\subsection{Black hole growth} \label{sec:black_holes}
Following \cite{DiMatteo2008} and \cite{BoothSchaye2009} we seed haloes with black holes (BHs) during the simulation. Starting at $z=19$ we run a friends of friends group finder every time the expansion factor increases by a factor $1.00751$. We seed a BH in every group that is above a certain mass threshold and that does not already have a BH. We seed BHs in haloes above a mass of $2.757 \times 10^{11} \  {\rm M}_{\odot} (m_\text{g} / 1.07 \times 10^9 \ {\rm M}_{\odot})$, corresponding to roughly fifty dark matter particles at each resolution. Because the \citet{BondiHoyle} accretion rate is proportional to the square of the BH mass, an increase in initial mass can cause BHs to grow much earlier. We use a BH seed mass of $10^5~{\rm M}_{\odot}$ for intermediate and high resolution, and of $10^7~{\rm M}_{\odot}$ for low resolution. The seed mass had to be increased for low resolution, since the rapid growth phase of the BHs corresponds to unresolved galaxy masses \citep[see e.g.][]{Bower2017,McAlpine2018}.

As we do not properly resolve dynamical friction at our resolution, BHs are repositioned by hand to the minimum of the gravitational potential following the method of \cite{bahe2021}\footnote{The exclusion of the BH from the calculation of the gravitational potential  used for repositioning was only done for high and low resolution, as we only became aware of its importance later. This significantly strengthened the quenching of star formation in galaxies with large stellar masses for our high resolution simulations.}. For BH mergers we also follow the prescription by \cite{bahe2021}.

Besides merging with other BHs, BHs grow via accretion of gas, which is assumed to occur at a modified Bondi-Hoyle rate,
\begin{equation}
    \dot m_{\rm accr} = \alpha\frac{4\pi G c^2 m^2_{\rm BH} \rho}{\left(c_{\rm s}^2 + v_{\rm BH}^2\right)^{3/2}},
\end{equation}
where $m_{\rm BH}$ is the BH mass, $c_{\rm s}$ is the sound speed of the gas, $\rho$ is the gas density, $c$ is the speed of light and $v_{\rm BH}$ is the velocity of the BH with respect to its environment. The factor $\alpha$ is a boost factor that is added because we do not resolve the Bondi radius and because we lack the resolution to model the phase structure of the ISM. We use the parametrization of \citet{BoothSchaye2009},
\begin{equation}
    \alpha = {\rm max}\left[\left(\frac{n_{\rm H}}{n_{{\rm H},*}}\right)^{\beta_{\text{BH}}},1\right],
\end{equation}
where $n_{{\rm H},*} = 0.1~{\rm cm}^{-3}$, which corresponds to the density threshold for star formation in the intermediate- and high-resolution simulations (we use the same value for all resolutions). The logarithmic density slope $\beta_{\text{BH}}$ is a free parameter that we vary during the calibration. After some experimentation using simulations where only a single parameter is varied between runs, we settled on priors of $0-0.9$, $0.1-0.9$ and $0-3$ for high , intermediate  and low resolution, respectively.

The gas accretion rate is capped at the \citet{Eddington1913} rate. Following \citet{bahe2021}, the BH is allowed to `nibble' on neighbouring gas particles until the gas particles only have half of their original mass remaining. 

\subsection{AGN feedback}\label{Sec:agn_feedback}
In all but two of the simulations AGN feedback energy is injected into the medium surrounding the BH in thermal form using the prescription from \citet{BoothSchaye2009}. The model used in the remaining simulations is based on jet feedback and is described in \S\ref{sec:jet_feedback}. 

While accreting gas, the BH adds a fraction $\epsilon_\text{r}\epsilon_\text{f}=0.015$ of the accreted rest mass energy to an internal feedback  energy reservoir, where $\epsilon_\text{r}=0.1$ is the assumed radiative efficiency and $\epsilon_\text{f}=0.15$ is the assumed AGN feedback efficiency, i.e.\ the fraction of the radiated energy that is coupled to the gas surrounding the BH. Once enough energy is available to increase the temperature of $n_{\rm heat}$ gas particles by $\Delta T_\text{AGN}$, this energy is injected into the neighbouring gas particles. The energy injected in a single event is proportional to $n_{\rm heat}\Delta T_{\rm AGN}$, 
where $\Delta T_{\rm AGN}$ is the increase in temperature that is applied to $n_{\rm heat}$ neighbours. We find that it is the product  $n_{\rm heat}\Delta T_{\rm AGN}$ that is most important for regulating how much gas is expelled from clusters, and that $\Delta T_{\rm AGN}$ and $n_{\rm heat}$ are largely degenerate. We therefore fix $n_{\rm heat}$ to one and use $\Delta T_{\rm AGN}$ as a free parameter that is varied in the calibration. Following the findings by \cite{Chaikin2022}, we inject the thermal energy into the nearest neighbour of the BH, which gives results that are nearly indistinguishable from a statistically isotropic approach.

To choose the prior for $\Delta T_\text{AGN}$ we take a similar approach as for the stellar feedback kick velocity. However, instead of avoiding velocities that are too low to have an effect, we now have to make sure that feedback raises the temperature to a value sufficiently high to avoid catastrophic numerical over-cooling. The sampling issue is also slightly different than for stellar feedback. While stellar feedback is limited to young stars, BHs can inject energy throughout their lives and hence the time sampling of these events becomes important. If the time between AGN feedback events becomes too long, then the BHs will be unable to self-regulate. If BHs cannot regulate their growth, then this can lead to an unrealistic mass distribution of both the BHs and their host galaxies. To summarise, we have two main considerations:
\begin{enumerate}
    \item What is the $\Delta T_\text{AGN}$ below which radiative losses are already severe at injection for the densities at which stars form? 
    \item What is the $\Delta T_\text{AGN}$ above which the time between AGN events becomes longer than the BH growth time?
\end{enumerate}

\citet{DVSchaye2012Therm} demonstrated that the density above which thermal feedback becomes ineffective can be predicted based on the ratio of the radiative cooling time, which depends on the density and temperature, and the sound crossing time across a resolution element, which depends on the numerical resolution. According to their equation~18, feedback becomes inefficient for densities exceeding
\begin{equation}
    n_{{\rm H} ,t_{c}} = 0.25 ~{\rm cm}^{-3} \left( \frac{\Delta T_{\rm AGN}}{10^{7.5} \ {\rm K}} \right)^{3/2} \left (\frac{m_\text{g}}{1.09\times 10^9\,\text{M}_\odot}\right )^{-1/2} \label{eq:DV&Sfornh}.
\end{equation}
Comparing this to our threshold for star formation ($n_{\rm H} = 10^{-1}\,\text{cm}^{-3}$ for intermediate/high resolution and $10^{-3}\,\text{cm}^{-3}$ for low resolution), yields minimum values of $\log_{10} \Delta T_{\rm AGN}/\text{K} = 6.9$, 7.2, and 6.2 for the high, intermediate, and low resolution, respectively. However, the above equation assumes radiative losses to be dominated by Bremsstrahlung and \citet{DVSchaye2012Therm} showed that it underestimates the radiative losses for $\Delta T_{\rm AGN} < 10^7\,\text{K}$. For this reason we do not consider values below $10^7\,$K. On the other hand, since we inject the energy at the end of the time step, the feedback can do work during a single time step even if the temperature is too low to avoid overcooling, which means that somewhat lower values than implied by the above equation (but still higher than $10^7\,$K) may still be of interest.

If we define $\Delta m_\text{BH}$ to be the gas mass that must be accreted for the BH to have sufficient energy to heat a single gas particle, then   
the ratio of the time between AGN feedback events and the time of BH growth is given by \citep{BoothSchaye2009},
\begin{align}
    \frac{t_\text{AGN}}{t_\text{BH}} &= \frac{\Delta m_\text{BH}/\dot{m}_\text{BH}}{m_\text{BH}/\dot{m}_\text{BH}} \\
    &= \frac{m_{\rm g}k_{\rm B} (1-\epsilon_\text{r})}{(\gamma - 1)\mu m_{\rm H}\epsilon_{\rm f} \epsilon_{\rm r} c^2} \frac{n_{\rm heat}\Delta T_{\rm AGN}}{m_\text{BH}}
\end{align}
\begin{equation}
    \begin{aligned}
        \qquad \ \ \approx 0.98 \left(\frac{1-\epsilon_{r}}{0.9}\right)\left(\frac{m_\text{g}}{1.09\times 10^9 \text{M}_\odot}\right)\left(\frac{\epsilon_{\rm f}\epsilon_{\rm r}}{0.015}\right)^{-1}\times \\  \left(\frac{n_{\rm heat}\Delta T_{\rm AGN}}{10^{8.5} \ {\rm K}} \right) \left (\frac{m_\text{BH}}{10^{7}\,\text{M}_\odot }\right )^{-1},
    \end{aligned}
\end{equation}
where $\gamma=5/3$ is the ratio of specific heats and $\mu=0.6$ is the mean particle mass in units of the proton mass $m_\text{H}$. Given that we expect to need AGN feedback to quench star formation in galaxies with stellar mass $M_* \gtrsim 10^{11}\,\text{M}_\odot$ and that in this mass range BHs are observed to have masses $M_\text{BH}\sim 10^{-3}\,M_*$ \citep{Haring2004}, we need the BHs to become self-regulating when $M_\text{BH} \ll 10^{8}\,\text{M}_\odot$. The condition $t_\text{AGN} < t_\text{BH}$ then implies that for our $n_\text{heat}=1$ we require $\Delta T_\text{AGN} \lesssim 10^{8.5}\,$K for intermediate resolution, and values 8 times higher (lower) for high (low) resolution. 

Based on the above considerations and some small test runs, we adopted the flat priors $\log_{10} \Delta T_{\rm AGN}/\text{K} =  7.7-8.9$, $7.5-8.5$, and $7.0-9.5$ for high, intermediate and low resolution, respectively. For both intermediate and high resolution the prior ranges are smaller than what is possible based on our considerations. From our test runs we found that these ranges bracket a sufficiently large range in the observables we are interested in and the smaller ranges lead to slightly better sampling of the parameter space around the best-fitting model. For low resolution the prior extends to (unnecessarily) high values, but we will see that the best-fitting value is actually similar to those for the other resolutions. We can afford a larger prior range for the low resolution simulations as we are only sampling two parameters.

\subsubsection{Jet feedback}\label{sec:jet_feedback}
In addition to the fully thermal AGN feedback scheme described above, we also calibrate a kinetic AGN feedback variation. The model used for kinetic AGN feedback is based on the spin-driven jet feedback model described by \cite{Husko2022}, implemented into \textsc{swift}. In this model energy is injected by kicking two particles on opposite sides of the BH, according to its angular momentum vector. The angular momentum of the BH is calculated in a subgrid model for an accretion disc that is based on general relativistic magneto-hydrodynamics simulations of single BHs in the low accretion regime (< 0.01 Eddington). For more details see \cite{Husko2022}. The spin from black holes that remains after mergers is computed according to the description by \cite{BHmerge2008}.

Due to the relatively low resolutions used for FLAMINGO, we make some simplifications to the complete model. As we intend for the jet model to be maximally different from the thermal feedback mode, we do not switch from kinetic to thermal feedback at high Eddington rates, and instead use the kinetic feedback at all accretion rates. Instead of using the efficiencies based on the subgrid accretion model, we fix the jet efficiency to $\epsilon=0.015$. This efficiency is equal to the combined coupling and radiative efficiency, $\epsilon_\text{f}\epsilon_\text{r}$, for the thermal mode feedback. This implies that for each unit of mass accreted by the BH, the same amount of energy becomes available in the jet model as for the fiducial thermal model. While we do not use a spin-dependent feedback efficiency, we do still use the subgrid model to track the angular momentum vector of the BH and use it to select the direction in which gas particles are kicked. The BH accretion model is identical to that described in \S\ref{sec:black_holes}, and for calibration of the jet model we vary the boost factor $\beta_{\text{BH}}$.

When the BH has accreted enough mass, two neighbouring gas particles are kicked with a total kinetic energy equal to
\begin{equation}
    E_{\text{jet}} = 2\times\frac{1}{2}m_{\rm g}v_{\text{jet}}^2,
\end{equation}
where $v_{\text{jet}}$ is the target jet velocity (we use the term target because it is the energy that is fixed, similarly to the supernova kicks, see \S\ref{sec:stellar_feedback}), which is a free parameter that we calibrate. The jet velocity plays a role similar to $\Delta T_{\text{AGN}}$ for the case of thermal feedback. As the energy is injected in kinetic form, the model is less affected by thermal losses, but picking velocities that are too low will make the gas unable to escape to large distances \citep[see][]{Husko2022}. For very high values we again run into sampling issues. Based on these considerations and some initial tests, we use flat priors over the range of $v_{\text{jet}} / (\text{km}~\text{s}^{-1}) = 10^{2.7}-10^{3.5}$, corresponding in energy to $\Delta T_{\rm AGN}/\rm{K} \approx 10^{7.1}-10^{8.7}$. We only calibrate this model at intermediate resolution.

\begin{figure*}
    \centering
    \includegraphics[width=\textwidth]{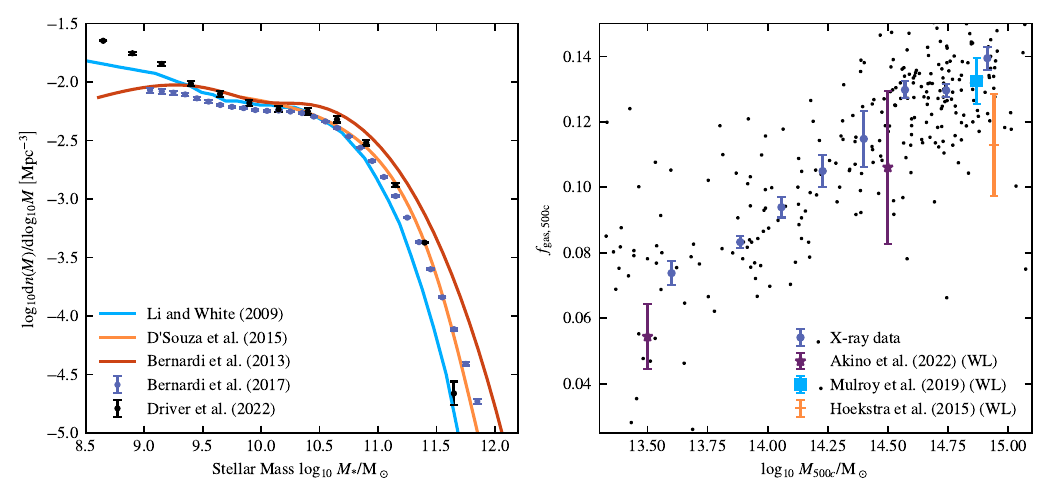}
    \caption{Compilation of observational data used for calibration. On the left we plot the SMF. On the right we plot the cluster gas fraction versus total mass, both measured at $R_{500c}$. Where available we display the $1\sigma$ measurement errors, which do not include intrinsic scatter. The X-ray data are binned from a compilation of available data, see \S\ref{sec:xraydata}, except the lowest mass point, which is obtained from a fit by \citet{Lovisari2015}. We show the individual clusters as black dots. Note that the X-ray data are plotted without any correction for the hydrostatic mass bias. For this work we use the \citet{Driver2022} data for the SMF, and the X-ray and \citet{Akino2022} data for the gas fractions.}
    \label{fig:all_data}
\end{figure*}

\begin{table*}
\caption{Mass ranges used for each observable when fitting the emulator to data. The values are rounded because the exact ranges vary with the values of the observational bias factors.}
\label{tab:massranges}
\begin{tabular}{lllll}
\hline Observable & SMF $M_*$ lower limit $\rm M_\odot$) & SMF $M_*$ upper limit $(\rm M_\odot$) & $f_\text{gas,500c}$ $M_{500c}$ lower limit $(\rm M_\odot$) & $f_\text{gas,500c}$ $M_{500c}$ upper limit $(\rm M_\odot$)  \\
    \hline High-res [m8] & $10^{8.67}$ & $10^{11.50}$ & $10^{13.50}$ & $10^{13.73}$ \\
    Intermediate-res [m9] & $10^{9.92}$ & $10^{11.50}$ & $10^{13.50}$ & $10^{14.36}$ \\
    Low-res [m10]         & $10^{11.17}$ & $10^{11.50}$ & $10^{13.50}$ & $10^{14.53}$ \\ \hline
\end{tabular}
\end{table*}

\section{Observational data and biases}\label{sec:obsdata}
Before we can start to calibrate our simulations, we need to have observational data to compare with our simulations. We calibrate to the galaxy stellar mass function (SMF) and the gas fractions in groups and clusters ($f_\text{gas,500c}(M_{500c})$).

One of the goals of the FLAMINGO simulations is to predict galaxy clustering and cross correlations between galaxies and other tracers of the matter distribution. The SMF allows us to constrain the stellar content of haloes as a function of their mass. This is not only crucial for the prediction of observations using galaxies, the stellar mass also directly affects the distribution of dark matter in haloes, and the orbits of subhaloes. Although matching the SMF does not ensure that each halo contains the correct stellar mass, it suggests the relation is at least statistically plausible provided the model assumes the correct cosmology. 

Besides galaxy clustering, we also wish to use FLAMINGO to investigate other cosmological observables tracing the distribution of matter, such as X-ray emission, the Sunyaev-Zeldovich (SZ) effect and lensing maps. From studies by \cite{Semboloni2013}, \cite{vanDaalen2020} and \cite{Salcido2023} we know that the gas fractions in clusters have a large impact on the matter power spectra on scales relevant for e.g.\ cosmic shear. By calibrating to the observed gas fractions, we can also make robust predictions for the distribution of gas expelled from group/cluster cores.

We calibrate to the same observables as were used for the BAHAMAS simulation \citep{BAHAMAS2017,BAHAMAS2018}. In this section we will discuss the data that we considered and the observational biases that we account for.

\subsection{The galaxy stellar mass function}\label{sec:obs_smf}
Constraining the SMF has been the goal of a large number of studies, many of which are based on the SDSS \citep{LiWhite2009,Dsouza2015,Bernardi2013,bernardi2017} or the more recent GAMA survey \citep{Baldry2012,Wright2017,Driver2022}. A compilation of these data sets is shown in the left panel of Fig.~\ref{fig:all_data}. It is clear that there are substantial systematic differences between some of the different groups that have tried to measure the SMF, particularly at the low- and high-mass ends. However, some of the most significant outliers are older results. While there are still discrepancies at the high-mass end, the results from the three most recent studies, \citet{Dsouza2015,bernardi2017,Driver2022}, are in reasonable agreement over a large part of the mass range. Instead of trying to combine different data sets, we limit the fitted mass range to $M_* < 10^{11.5}\,\text{M}_\odot$ and we choose to use the most recent GAMA result from \citet{Driver2022} at $z=0$. Not only is this the most recent study, it also provides a useful prior for possible biasing due to cosmic variance. The upper mass limit also decreases the possible bias we get due to our choice of simulation aperture (see \S\ref{sec:sim_output} and Appendix~\ref{sec:ap_aper} for more details). We always set a simulation-resolution dependent lower mass limit on the mass range we use for fitting. The mass ranges we use can be found in Table~\ref{tab:massranges}.

Fitting the SMFs from simulations to observations requires special care. There are some important differences/sources of uncertainty that need to be taken into account:
\begin{enumerate}
    \item Observations suffer from random errors in measuring the mass. while simulations have no mass measurement errors (at least for a fixed definition of a galaxy, i.e.\ for a given subhalo finder). Simulations do suffer from randomness errors \citep[see][]{Borrow2022}, as discussed by these authors, this issue is negligible for our analysis because we consider large ensembles of galaxies..
    \item Observations possibly suffer from systematic errors, which may originate from spectral energy distribution fitting, corrections for dust extinction, surface brightness profile fitting, and/or selection effects.
    \item Observations may suffer from cosmic variance.
\end{enumerate}
Before discussing how we take each of these effects into account, we note that the uncertainty in the stellar IMF is not directly relevant because the observational analysis and the simulations use the same IMF. The observed SMF also depends on the assumed cosmology, but this is close enough to the one used in the simulations to have a negligible effect on the comparison. 

\subsubsection{Random errors on the observed stellar mass}\label{sec:eddbias}
Symmetric observational scatter in the measured stellar mass will cause a systematic shift in the inferred SMF. Because there are more galaxies in lower mass bins, it is more likely for galaxies to scatter to a higher mass bin than to a lower mass bin. This is especially important at the high-mass end, where the SMF is steep. This effect is known as \cite{Eddington1913} bias. We account for it by adding scatter to the simulation masses. We adopt the lognormal scatter from \cite{Behroozi2019}, which has a redshift-dependent standard deviation of
\begin{equation}
       \sigma(\log_{10}M_{\ast}) =  {\rm min}\left(0.070+0.071z, 0.3\right) \ {\rm dex},
\end{equation}
where we sample the lognormal distribution for each galaxy. This then adds an Eddington-like bias to the simulation results, consistent with observations.

\subsubsection{Systematic errors in the observed stellar mass} \label{sec:mstar_sysbias}
There are systematic discrepancies between the different observations. The reason for this is mostly found in the stellar population synthesis and dust correction models used, as the observed luminosity functions agree better between different studies than the mass functions. However, at the FLAMINGO resolution, the stellar masses can be predicted much more accurately than the star formation histories, current-day star formation rates and dust extinction rates. Therefore, calibration to the SMF is preferable over a direct comparison with the luminosity function.

To account for potential systematic shifts in the observed stellar masses, we include a stellar mass bias parameter
\begin{equation}
    \log_{10}(M_{*,\rm obs}) \rightarrow 
   \log_{10}(M_{*,\rm obs}) + \log_{10} b_*,
\end{equation}
where the bias $b_*$ is assumed to be independent of mass. Note that the sign is defined such that a positive stellar mass bias implies the observations underestimate the true stellar mass. We use a lognormal prior to constrain the bias parameter. The prior is taken from ~~~\cite{Behroozi2019} (their eq.~25) and is based on the existing tensions between observed time-integrated star formation rates and observed SMFs,
\begin{equation}
    \log_{10} b_* = \mathcal{N}(0,0.14),
\end{equation}
where $\mathcal{N}(\mu,\sigma)$ is a normal distribution with mean $\mu$ and standard deviation $\sigma$. 

We adopt a mass-independent bias. While a mass-dependent bias might have improved the agreement between the data and the simulations, the mass dependence is unknown and therefore there is no obvious parametrization of the mass dependence. This implies the new free parameters would have no clear priors. Additionally, we note that our decision not to fit above a stellar mass of $10^{11.5}~\rm{M}_{\odot}$ has a similar effect as switching to a much higher stellar mass bias above this mass.

\subsubsection{Cosmic variance}\label{sec:cosmic_var}
\citet{DriverCV2010} showed that the error on the SMF due to cosmic variance can be $5-10$ per cent for surveys like GAMA and the SDSS, depending on the volume considered. Cosmic variance can bias the number density measurements, because the survey may consist of slightly over- or under-dense regions. For our mass range we assume that this effect is independent of mass (S.~P.~Driver,~private~communication). To account for cosmic variance, we allow the observed number densities to shift up and down slightly,
\begin{equation}
    f_\text{obs} \rightarrow f_\text{obs} + \log_{10}(b_\text{cv}).
\end{equation}
Note that the sign is defined such that a positive cosmic variance bias implies the observations underestimate the number density of galaxies.
We constrain this bias parameter with a Gaussian prior taken from \cite{Driver2022}. They estimate the error due to cosmic variance to be about 6 per cent, so our prior is given by
\begin{equation}
    b_\text{cv} = \mathcal{N}(1,0.06).
\end{equation}

\begin{table*}
    \caption{Overview of the cluster gas mass fraction data used for this work. The first column lists the reference from which the data were obtained, the second column lists the number of objects, where 'fit' indicates that the main result is a fitted relation between $M_{500c}$ and $f_{\text{gas},500c}$, the third column shows how the total mass was measured (HSE: X-ray data assuming hydrostatic equilibrium; WL: weak gravitational lensing), and the final column contains comments on the selection method.}
    \vspace{5pt}
    \begin{tabular}{l|r|l|l}
    \hline Reference  & $N$ & Type & Selection \\
     \hline \cite{Vikhlinin2006} & 10 & HSE & Nearby, relaxed, ambiguous X-ray limit \\
     \cite{Maughan2008} & 114 & HSE & NED Cross-match, $z>0.1$
     \\ \cite{Rasmusses2009} & 15 & HSE & Bright groups
    \\ \cite{Sun2009} & 23 & HSE &$0.015<z<0.13$, resolved temperature profiles \\
    \cite{Pratt2010} & 31 & HSE & X-ray flux limited, $z < 0.2$ \\
    \cite{Lin2012} & 94 & HSE & Infrared magnitude limited \\
    \cite{Lagana2013} & 126 & HSE & Crossmatch between \cite{Maughan2008} and SDSS; X-ray flux limit \\
    \cite{Sanderson2013} & 5 & HSE & Optical magnitude limit, $\sigma\leq500c$ km~s$^{-1}$ \\
    \cite{Gonzalez2013} & 15 & HSE & Optical magnitude limit, $0.03<z<0.13$ \\
    \cite{Lovisari2015} & 20 & HSE & X-ray flux limited \\
    \cite{Hoekstra2015} & 50 & WL & X-ray flux limited\\
    \cite{Pearson2017} & 8 & HSE & GAMA r-band selection, $N>12$, $z<0.12$ \\
    \cite{Mulroy2019} & fit & WL & X-ray luminosity limit \\
    \cite{Lovisari2020} & 120 & HSE & tSZ-selected from Planck data. \\
    \cite{Akino2022} & fit & WL & C1 - X-ray selected, C2 no clear selection.\\
    \hline
    \end{tabular}
    \label{tab:all_data}
\end{table*}

\subsection{The cluster gas mass fractions}\label{sec:obs_fgas}
Data for the cluster gas mass fractions, $f_\text{gas,500c}$, come in two varieties. They are either obtained purely from X-ray observations, or from a combination of X-ray and weak gravitational lensing observations where the latter are used to measure the total cluster mass. For the X-ray only data, the density and temperature profiles fitted to the observations are used to measure the total mass assuming the gas is in hydrostatic equilibrium (HSE). In both cases the gas mass is obtained by integrating the density profile measured from X-ray observations out to the measured value of $R_{500c}$. Table~\ref{tab:all_data} summarises all the different sets of data that we use. 

As was the case for the SMF, there are biases that we need to account for when we compare observations with simulations.
There are four distinct issues that we take into account:
\begin{enumerate}
    \item At the low-mass end selection effects become important, because at fixed halo mass objects with a higher gas content will tend to emit more X-ray radiation. Any X-ray selected sample may therefore have gas fractions that are biased high, particularly at low masses.
    \item The measurement of total mass from X-ray data under the assumption of HSE is well documented to be biased low \citep[e.g.][]{Hoekstra2015,Eckert2016,Smith2016}.
    \item For the weak lensing data, we make use of the fits of the relation between gas fraction and mass provided by the authors. The fits are preferred to individual measurements as the fits account for the selection function of the sample. However, for our purposes the fits need to be sampled at particular masses. This needs to be done in a way that limits the covariance between the samples and that is representative of the data used (i.e.\ no extrapolation).
    \item As clusters are rare objects they are usually observed over a large redshift range. Furthermore, because weak lensing is most efficient when the lens is halfway between the observer and the background galaxies, weak lensing observations tend to probe higher redshifts than X-ray data. Clusters evolve over time, so we need to make sure the simulation samples are representative for the observational samples we compare them with.
\end{enumerate}
For the cluster gas fractions the largest mass we can fit for is limited by the box size of each simulation. The upper mass limit used for fitting therefore changes with resolution (as we use a different box size for each resolution). The upper limits can be found in Table~\ref{tab:massranges}.

\subsubsection{X-ray data}\label{sec:xraydata}
The first set of gas fraction data we describe is the X-ray (or HSE) data. For each data set we store $M_{500c}$ and $f_{\rm{gas}, 500c}$, with asymmetric errors where available, and correct the data to the FLAMINGO cosmology ($M_{500c} \propto h^{-1}$, $f_{\rm{gas}, 500c} \propto h^{-1.5}$).
The combined data set has 581 objects but contains duplicates. For each object that appears more than once we calculate a new data point by taking an unweighted mean of the different measurements. The mean is taken in both $M_{500c}$ and $f_{\rm{gas}, 500c}$. Because the duplicates are often based on (in part) the same data, the errors will not be independent and we combine them via 
\begin{equation}
    \sigma^2 = \frac{1}{N}\sum_{i}^N\sigma_i^2,
\end{equation}
where $N$ is the number of times a single object appears in the set. This leaves us with 533 objects. Note that we do not use the errors for the re-binning, as we make use of bootstrap re-sampling to compute the errors.

We need to consider redshift evolution. The emulators will be trained on simulation snapshots corresponding to a single redshift. Imposing a redshift cut of $z<0.25$ causes the median redshift of the X-ray sample to become 0.1, thus allowing us to compare with simulation snapshots at $z = 0.1$. The redshift cut reduces the sample to 310 objects. The individual masses and gas fractions are shown as black dots in Fig.~\ref{fig:all_data}.

\begin{table}
    \caption{Compilation of cluster X-ray gas fraction data used for calibration. These values are for the DESYR3 cosmology ($h=0.681$, $\Omega_\text{m}$ = 0.298). The values are obtained by taking the median of the X-ray data described in Table~\ref{tab:all_data} in eight logarithmically spaced bins between $10^{13.8}$ and $10^{15.0}~\text{M}_{\odot}$. The errors are the absolute difference between the 16th or 84th percentile and the median (whichever is largest), obtained by bootstrap resampling the median.}
    \label{tab:HSE_vals}
    \centering
\begin{tabular}{l|l|l|l}
\hline
 $M_{500c}$ & $f_{\text{gas},500c}$ \\
 ($\log_{10}{\rm M}_{\odot}$) & \\
  \hline 13.89 & $0.083\pm 0.002$\\
   14.06 & $0.094\pm 0.003$\\
   14.23 & $0.105\pm 0.005$\\
   14.40 & $0.115\pm 0.008$\\
   14.57 & $0.130\pm 0.002$\\
   14.74 & $0.130\pm 0.002$\\
   14.91 & $0.139\pm 0.003$\\
   \hline
\end{tabular}
\end{table}

We combine the X-ray measurements by computing the median gas fraction in eight logarithmically spaced hydrostatic mass bins between $10^{13.8}$ and $10^{15.0}~\text{M}_{\odot}$. For each bin, the error on the median is obtained by taking the difference between the median and the 16th$-84$th percentiles obtained from bootstrap resampling the objects. This gives us asymmetric errors around the median. As our likelihood uses symmetric errors, we use only the greater of the positive and negative errors. The tabulated data points can be found in Table~\ref{tab:HSE_vals}.

Furthermore, selection effects are expected to be most prevalent at lower halo masses. The median observed gas fraction as a function of mass shows a clear trend-break at $M_{500c,\rm{HSE}}\approx10^{13.8}~{\rm M}_{\odot}$. Below this mass the gas fractions no longer decrease, but instead plateau, a behaviour that deviates from what is expected for an unbiased sample \citep[e.g.][]{BAHAMAS2017}. To deal with this we impose a mass cut at a hydrostatic mass of $M_\text{500c,HSE} > 10^{13.8}~{\rm M}_{\odot}$, but add the fit from \cite{Lovisari2015} at their median mass ($4\times 10^{13}\,\text{M}_\odot$) as a separate data point.

We account for hydrostatic mass bias by adding a constant bias term to the HSE masses,
\begin{equation}
    \log_{10} M_{500c}  = \log_{10}M_{500c, \rm{HSE}} - \log_{10}(b_{\rm HSE}).
\end{equation}
Note that values $b_\text{HSE}<1$ imply that the hydrostatic mass estimate underestimates the true mass. We neglect the effect of hydrostatic bias on the gas fraction because it is comparatively small \citep{BAHAMAS2017}. This is because both the total and gas mass increase with increasing $R_{500\rm{c}}$. The measured gas fraction will differ only at the level of the change in cumulative gas fraction between the true and biased $R_{500\rm{c}}$. This is expected to cause only mild changes in the gas fraction \citep[see e.g.\ fig.~6 of][]{Velliscig2014}. Before calculating the median that we compare with the simulation we thus adjust all the observed HSE masses. By combining both X-ray and weak lensing observations, we can constrain the hydrostatic bias. However, we found that our compilation of data on its own is not constraining enough without the use of a prior. To define our prior, we take the values $0.72\pm0.08$ from \citet{Eckert2016} and $0.76\pm0.06$ from \citet{Hoekstra2015} and combine the two to obtain the Gaussian prior
\begin{equation}
    b_{\rm HSE} = \mathcal{N}(0.74,0.10).
\end{equation}
\citet{Eckert2016} and \citet{Hoekstra2015} estimate the hydrostatic mass bias by directly comparing the masses they obtain from weak lensing and from X-rays.

\subsubsection{Weak lensing data}
We complement the X-ray data with the latest HSC-XXL weak gravitational lensing data from \cite{Akino2022}. Higher-mass data from \cite{Mulroy2019} and \cite{Hoekstra2015} are available and plotted in Fig.~\ref{fig:all_data}, but the box size used for our calibration runs is too small to make use of them. To compare with the weak lensing data, we make use of the power-law fits to the relation between the gas fraction and mass given by the authors. These fits take selection effects into account. Because the power-law fits have two free parameters, sampling them at more than two masses would result in strong covariance between the sampled points. We therefore use the fit to create two data points that are spaced equally far from the pivot used by the authors. This gives us $f_{\text{gas},500c}(M_{500c}=10^{13.5}~\text{M}_{\odot})=0.054\pm0.010$ and $f_{\text{gas},500c}(M_{500c}=10^{14.5}~\text{M}_{\odot})=0.106\pm0.023$. Due to the limited box size, we use only the lower, $M_{500c}=10^{13.5}~{\rm M}_{\odot}$, point for fitting high- and intermediate-resolution simulations. For low resolution we are able to include the second $M_{500c}=10^{14.5}\,{\rm M}_{\odot}$ point.

The median redshift of the HSC-XXL sample is $z = 0.3$. We therefore construct a separate emulator for $f_\text{gas,500c}$ at $z=0.3$, which we use to fit the weak lensing data. The fits make use of self-similar scaling to move the different clusters to the same redshift, so we could have corrected them to the redshift $z=0.1$ used for the X-ray data. However, we prefer to use a redshift close to that of the actual sample, to minimize the size of the correction. \citet{Akino2022} give both the weak lensing inferred and the true $M_{500c}$, as they correct for the expected bias on the weak lensing inferred $M_{500c}$. We make use of their calibrated true $M_{500c}$ masses.

\section{Emulator construction} \label{sec:emulator_descr}
Cosmological hydrodynamical simulations are too expensive to be run for each step in an MCMC chain used to evaluate likelihoods. In order to use simulation outputs in MCMC methods, we therefore make use of emulators trained on a set of simulations. Emulators are used to interpolate results in the parameter space between training simulations. They are able to predict the output of the simulations as a continuous function of the input parameters, in a fraction of the original computation time. This method has previously been applied to the matter power spectrum \citep[e.g.][]{Heitmann2009a,Heitmann2016,EUCLIDemu,Angulo2021} and to baryonic observables \citep[e.g.][]{HydroMLCal1,HydroMLCal2}. By using emulators, we can interpolate between the results of a set of training simulations and obtain a fully continuous prediction of how the simulation responds to changes in subgrid parameters.

\subsection{Training sets}
The first step in setting up the emulator is to create a training set. In our training set we want to vary those subgrid parameters that we know are important for the calibration. As discussed in Section~\ref{sec:simulations}, for the intermediate- and high-resolution simulations we vary the following four parameters: the stellar feedback efficiency, $f_\text{SN}$, the target kick velocity for stellar feedback, $\Delta v_\text{SN}$, the power-law slope of the density dependence of the black hole accretion boost factor, $\beta_{\text{BH}}$, and the AGN heating temperature, $\Delta T_\text{AGN}$ ($v_{\rm jet}$, the target kick velocity for AGN feedback in the jet model). For the low-resolution simulations we do not require stellar feedback and therefore vary only the last two parameters. The ranges over which the parameters are varied are motivated in Section~\ref{sec:simulations} and listed in Table~\ref{tab:paramtab} (Table~\ref{tab:jet_params} for the jet model). 

To optimise the parameter space, we make use of a Latin hypercube, first proposed by \citet{LatinHypercube}. To set up a Latin hypercube with $N_\text{sims}$ nodes, we start with an ordered list of $N_\text{sims}$ independent samples along every dimension of the hypercube, where the number of dimensions equals the number of subgrid parameters that are varied. These samples are then combined and shuffled to create a set of $N_\text{sims}$ points $\boldsymbol{\theta}$ that are distributed uniformly within the hypercube, where in our case $\boldsymbol{\theta} = (f_\text{SN},\log_{10} \Delta v_\text{SN},\beta_{\text{BH}},\log_{10} \Delta T_\text{AGN})$ for intermediate and high resolution, and $\boldsymbol{\theta} = (\beta_{\text{BH}},\log_{10} \Delta T_\text{AGN})$ for low resolution. Our criterion for optimising the sampling is the 'maximin' approach, which maximises the minimum distance that sampled points are away from each other. An in depth explanation of how the method works is provided by \cite{Heitmann2009a}. We apply to each sample a random shift of at most half the average spacing between samples. We then run the $N_\text{sims}$ simulations corresponding to the nodes of the Latin hypercube.

We use the public package \textsc{swiftemulator}\footnote{\url{https://swiftemulator.readthedocs.io/en/latest/}} \citep{Kugel2022}, built on the package \textsc{george} \citep{george}, to set up the Latin hypercube as well as to train and test the emulators. \textsc{swiftemulator} streamlines the emulation process for results obtained from \swift~runs. Within \textsc{swiftemulator} we use the Latin hypercube generator from \textsc{pyDOE} \citep{pyDOE2012}.

\begin{figure}
    \centering
    \includegraphics[width=0.5\textwidth]{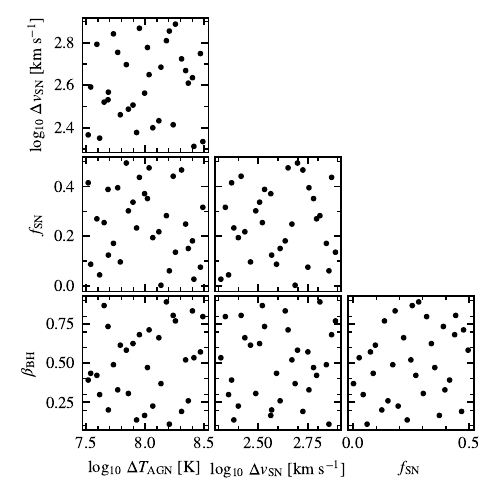}
    \caption{The sampling of parameters in the 32-node Latin hypercube used to train the emulator for the intermediate-resolution simulations.}
    \label{fig:hypercube_final}
\end{figure}

We use $N_\text{sims}=32$. The sampling of parameter space provided by the Latin hypercube used for intermediate resolution is shown in Fig.~\ref{fig:hypercube_final}. The box sizes used for the training are $(100 ~\rm Mpc)^3$, $(200 ~\rm Mpc)^3$ and $(400 ~\rm Mpc)^3$ for high, intermediate, and low resolution, respectively. The volume is a compromise between computational cost and the maximum mass for which we train the emulator. Each run cost $\sim 800$, $\sim 1300$ and $\sim 1600$ cpu hours for low, intermediate and high resolution respectively. Using single simulations with an eight times larger volume at each resolution and with the results of \citet{Flamingomain}, we have verified that these box sizes are sufficiently large for box size effects to be negligible with respect to the production runs.

\subsection{Obtaining the required simulation output} \label{sec:sim_output}
From our simulation we take three snapshots at $z=0$, 0.1 and 0.3. For each snapshot we find haloes and subhaloes using \textsc{VELOCIraptor} \citep{VR1,VR2}. After an initial friends of friends group search it uses the full 6-D phase space information to disentangle the central and satellite subhaloes. 

One of the difficulties of comparing with data, is that we have to choose how to define the edge of simulated galaxies. Observed cluster gas mass fractions are measured within $R_{500c}$. For the stellar masses needed to compute the SMF, the situation is less clear. Ideally, we would create mock observations, fit them with S\'{e}rsic profiles and integrate these to obtain stellar masses, which is the procedure adopted by observational studies. This was recently done for the EAGLE simulation by \cite{deGraaff2022}. However, the resolution of the FLAMINGO simulations is too limited to mimic the observational strategy. As shown by \citet{Flamingomain}, FLAMINGO significantly overestimates the sizes of low-intermediate mass galaxies, which means we cannot create realistic virtual galaxy observations. Based on the findings of \cite{deGraaff2022}, we choose to calibrate the SMF using a 3D aperture with a radius of 50~kpc for the simulations. A comparison between different choices of aperture can be found in Appendix~\ref{sec:ap_aper}, where we show that the aperture becomes only important above a stellar mass of $\approx 10^{11} {\rm M}_{\odot}$.

Before computing the galaxy SMF, we first add random errors to the simulation stellar masses as described in \S\ref{sec:eddbias}. The SMF is then sampled in 25 logarithmically spaced mass bins between $10^{9} \, {\rm M}_{\odot}$ and $2\times10^{12}  \, {\rm M}_{\odot}$ for intermediate- and low-resolution simulations, and 40 bins between $10^{8} \, {\rm M}_{\odot}$ and $2\times10^{13} \, {\rm M}_{\odot}$ for high-resolution simulations. We choose to use a finer binning than is available for the observational data to allow the emulator to capture the finer features of the predicted SMF. Tests with different binning strategies show this had no effect on the results. We have enough galaxies across the fitted mass range for the Poisson errors to still be very small even with finer binning. The uncertainty we provide to the emulator is the Poisson error for each bin.

For the gas fraction we instead opt for an adaptive binning strategy. While the simulation volumes used for the calibration are large enough to constrain the SMF over the adopted mass range, at the high cluster mass end, we always run out of clusters before we run out of data to compare with. For all resolutions we use 20 bins between $M_{500c}$ of $10^{13}$ and $10^{15}\, {\rm M}_{\odot}$ although we never manage to make use of this entire range. As the higher mass bins start to run out of objects, we allow the highest mass bin to stretch to include a sufficient number of objects. We require each bin to contain at least ten objects. We also limit the stretching of the bin to half the original bin width. The uncertainties we provide to the emulator are based on the 16th$-84$th percentiles. As the emulator only takes symmetrical errors, we take mean of the absolute difference between the median and 16th percentile and the difference between the median and 84th percentile. For both the SMF and the cluster gas fraction we discard any empty bins.

\subsection{Training using Gaussian processes}
After measuring the SMF and cluster gas fraction for each node of the hypercube, we can train an emulator for each observable. Because each individual node of the Latin hypercube requires a cosmological hydro simulation, we are operating in a regime where we have a limited number of samples. We also know a priori that the observables we want to emulate (i.e., the galaxy number density and group and cluster gas fractions) vary smoothly with mass and with the values of the subgrid parameters. Both these properties are in the regime in which Gaussian processes give excellent predictive power with respect to\ the input data \citep[see e.g.][]{Rasmussen2004,GPbook}. 

We set up a different Gaussian process for each relation we emulate. We combine the mass (either stellar or $M_{500c}$) and subgrid parameters into a single input data vector $\mathbfit{x}=(\log_{10} M,\boldsymbol{\theta})$, from which the emulator then predicts the dependent quantity, which is either the number density of galaxies, $f(M_*)$, or the gas fraction, $f_{\rm gas,500c}$. Each emulator thus has $N+1$ parameters, where $N$ is the number of subgrid parameters that are varied. In order to limit the dynamic range, we transformed many of the inputs to log-space. This includes the masses (aperture stellar mass or $M_{500c}$), the values of the SMF and the two subgrid parameters that are sampled in log-space ($\Delta v_\text{SN}$ and $\Delta T_\text{AGN}$). This is an important step as it greatly increases the smoothness of the emulated relations, making it much easier for the emulator to give accurate predictions. As the input relations are smooth over the range we are interested in, we do not require any other transformations of the input. We feed the data directly into the Gaussian process. We use a squared exponential kernel
\begin{equation}
    k(\mathbfit{x},\mathbfit{x}') = \exp\left(-\frac{(\mathbfit{x}-\mathbfit{x}')^{T}\mathbf{\Theta}^{-1}(\mathbfit{x}-\mathbfit{x}')}{2}\right),
\end{equation}
where $\mathbf{\Theta}$ represents a diagonal matrix containing the hyperparameters that set the scale for each input parameter, and $\mathbfit{x}$ and $\mathbfit{x'}$ are two positions in parameter space. The hyperparameters are optimised based on maximising the marginal likelihood \citep[see][]{GPbook}. As we train a separate Gaussian process for each relation, we also have a separate set of hyperparameters for each relation. We have verified the posteriors of the hyperparameters to ensure that the values we use are well converged.

\subsection{Error estimation} \label{sec:emu_accu}
It is important to verify that the emulator is able to give accurate results before we use it to find best-fitting subgrid and bias parameters. Moreover, we need to quantify the accuracy of the emulator because we will account for emulation errors when fitting to data. The best way to measure the uncertainty in the emulator predictions is to perform test simulations that span the emulated parameter space. However, this implies that we would need to run many additional simulations. To save time, we choose instead to measure the uncertainty by making use of k-fold cross-validation, which we will refer to as cross-checks.

\begin{figure*}
    \centering
    \includegraphics[width=\textwidth]{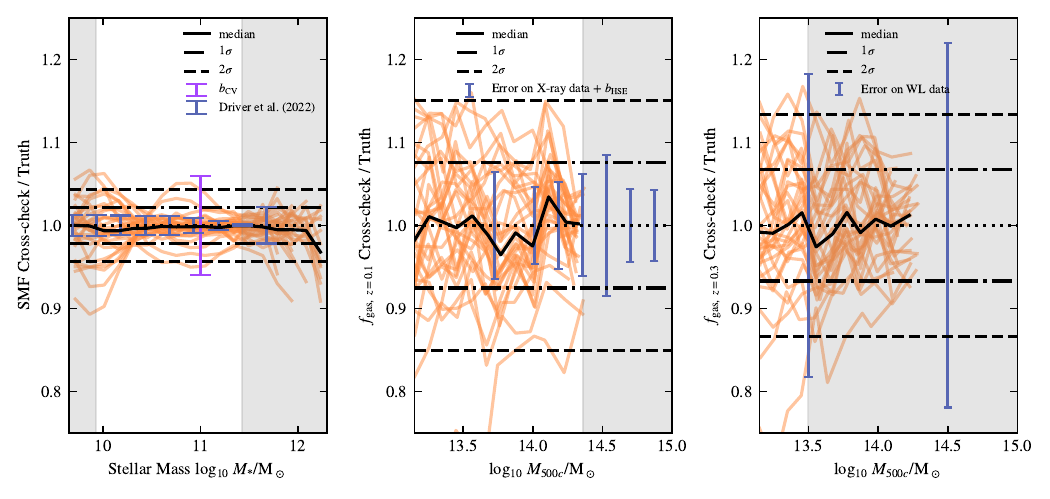}
    \caption{Performance of the emulator on cross checks (see \S\ref{sec:emu_accu}) for the redshift $z=0$ SMF (left panel), the $z=0.1$ X-ray cluster gas fractions (middle panel), and the $z=0.3$ weak lensing cluster gas fractions (right panel) at intermediate [m9] resolution. Each of the 32 red lines corresponds to the case where a single simulation from the 32-node Latin Hypercube has been omitted from the training set. The curves show the ratio of the emulator prediction for the parameter values of the omitted simulation to the actual simulation values. The solid black line shows the median as a function of mass. The horizontal dash-dotted and dashed lines indicate, respectively, the 1 and 2 $\sigma$ mean errors on the emulator. The horizontal dotted lines indicate the one-to-one lines, i.e.\ zero errors. The grey bands indicate the regions that are not used for fitting in Section~\ref{sec:likelihoods}. In each panel we also indicate the observational errors. For the SMF we show the error due to cosmic variance and the errors on the data by \citet{Driver2022}, for the $z=0.1$ gas fractions we combine the error from the X-ray data with the error due to hydrostatic bias and for the z=0.3 gas fraction we show the error on the weak lensing data by \citet{Akino2022}. The emulator predictions are accurate enough to predict to simulation output within the observed constraints}
    \label{fig:hyper_perform}
\end{figure*}

We create $N_{\rm sims}$ new data sets, where $N_{\rm sims}$ is the number of nodes in our Latin hypercube (32 in our case). For each of these data sets we take out one simulation and retrain the emulator on the reduced set of $N_\text{sims}-1$ samples. We then test how accurately the emulator is able to predict the simulation that was left out. We do this by taking the ratio between the result from the run that was left out, and the prediction of the emulator for the parameter values of the left-out run. This gives us a value for each mass bin in the training data. We combine the ratios for all mass bins and  $N_{\rm sims}$ emulators into a single list and compute the standard deviation, $\sigma_{\rm crosscheck}$. The error on the emulator prediction, $\sigma_{\rm emu}$, is then given by 
\begin{equation}\label{eq:emu_err}
    \sigma_{\rm emu} = \left|\sigma_{\rm crosscheck}f(M,\mathbf{\theta})\right|,
\end{equation}
where $f(M,\mathbf{\theta})$ is the value predicted by the emulator for mass $M$ and at parameter values $\mathbf{\theta}$. The result of the cross checks for the Latin hypercube of intermediate-resolution simulations can be seen in Fig.~\ref{fig:hyper_perform}. It is important to note that cross checks are a conservative method to estimate the uncertainty. The input for cross-checks is uniformly sampled, implying that a significant fraction of the test points is located near the boundaries of the parameter space, where a Gaussian process is naturally less accurate.

From Fig.~\ref{fig:hyper_perform} it is clear that our emulators do not suffer from significant systematic errors for our three calibration targets, the $z=0$ SMF, $z=0.1$ X-ray cluster gas fractions, and $z=0.3$ weak lensing cluster gas fractions. There are no significant trends with mass, and the medians ratio is centered close to one, which corresponds to an error of zero.

It is clear that the emulator for the SMF is more accurate than the emulators for the gas fractions. This is a reflection of the way we constrain the input simulations. In the case of the SMF, the errors on the input are Poisson errors, which are quite small for our simulation volumes in the mass range we are interested in. The $f_\text{gas}$ errors are based on the 16th$-84$th percentiles of the simulated gas fractions in each mass bin, which can be larger than the 5 per cent accuracy that the emulator attains.

The emulator accuracy for all resolutions can be found in Table~\ref{tab:acc_table}. The emulators become more accurate going to lower resolution. There are several possible reasons for this trend. First, we used larger box sizes for the lower-resolution simulations, so the uncertainty intrinsic to the simulation is smaller at fixed mass. Second, we used a slightly larger parameter range for high resolution than for intermediate resolution, while for low resolution we only used two parameters, greatly reducing the sampled space. 

\begin{table}
    \caption{Accuracy of the emulators, $\sigma_{\rm crosscheck}$, for the three different simulation resolutions and the jet model AGN variation, in percentages. The values are obtained by taking the standard deviation of the ratio between the result from the simulation omitted from the Latin hypercube and the prediction from the emulator trained on all but that simulation.}
    \centering
    \begin{tabular}{l|l l l l}
    \hline
    Calibration target & High & Intermediate & Low & Jet\\
    \hline $\log_{10}$ SMF & 2.7  & 2.2  & 1.5  & 1.9  \\
    $f_{\text{gas},z=0.1}$ & 8.9  & 7.5  & 4.8  & 7.1 \\
    $f_{\text{gas},z=0.3}$ & 7.9  & 6.7  & 4.2  & 6.1 \\
    \hline
    \end{tabular}
    \label{tab:acc_table}
\end{table}

The obtained accuracy is sufficient, as it is higher than the observational scatter/uncertainty. Any deviations between the model and the data at the level of the emulator error would still be consistent with the observational constraints, especially as we allow for observational biases in our analysis.

\section{Using the emulator for parameter estimation}\label{sec:likelihoods}
To use the emulator as the model that we compare with observational data, we need a way to optimise the subgrid parameters $\boldsymbol{\theta}$ (see Section~\ref{sec:simulations}) and, optionally, the observational bias factors $\log_{10}b_*$, $b_\text{CV}$, and $b_\text{HSE}$ (see Section~\ref{sec:obsdata}). 

For parameter optimisation we use the Markov chain Monte Carlo (MCMC) package \textsc{emcee} \citep{emcee}. We use the ensemble sampler, which we give our posterior likelihood. For every fit we have done using MCMC, we have varied the number of walkers and steps to ensure the resulting values are converged. We discard the first 500 steps of each chain to avoid systematic errors due to the burn-in phase.

To evaluate the goodness of fit of an emulator prediction to the observations, we first define the log likelihood for a single observed mass bin. For the SMF this is given by
\begin{multline}
    \ln \mathcal{P}_{\rm SMF}(M_{*,{\rm obs}},b_\text{cv},b_*,\boldsymbol{\theta}) \equiv  \\
    - \frac{\left [ f_\text{obs}(M_{*,{\rm obs}}) + \log_{10}b_{\rm CV} - f_\text{emu}(b_{*}M_{*,{\rm obs}},\boldsymbol{\theta})\right ]^2}{\sigma^2_{\rm obs}(M_{*,{\rm obs}}) + \sigma^2_{\rm emu}(b_{*}M_{*,{\rm obs}},\boldsymbol{\theta})},
\label{eq:P_SMF}
\end{multline}
Here $f(M_*)$ is the SMF, 
\begin{equation}
    f(M_*) \equiv \log_{10} \left (\frac{\dd n}{\dd\log_{10}(M_*)} \right ),
\end{equation}
the subscripts indicate whether the quantity is observed ('obs') or emulated ('emu'), $\boldsymbol{\theta}$ is a vector containing the values of the varied subgrid parameters, and $\sigma$ is the error on $f$. For $\sigma_{\rm emu}$ this refers to the error on the emulator from cross-checks, equation~\ref{eq:emu_err}. The expression also accounts for observational bias factors due to cosmic variance, $b_\text{CV}$, and the conversion of direct observables into stellar mass, $b_*$, that were discussed in \S\ref{sec:obs_smf}. For cluster gas fractions measured from X-ray observations the log likelihood is defined as
\begin{multline}
    \ln \mathcal{P}_\text{gas}(M_{500c,{\rm obs}},b_\text{HSE},\boldsymbol{\theta}) \equiv \\
    - \frac{\left [f_\text{gas,500c,obs}(M_{500c,{\rm obs}}) - f_\text{gas,500c,emu}(b_\text{HSE}^{-1}M_\text{500c,obs},\boldsymbol{\theta})\right ]^2}{\sigma^2_{\rm obs}(M_{500c,{\rm obs}}) + \sigma^2_{\rm emu}(b_\text{HSE}^{-1}M_\text{500c,obs},\boldsymbol{\theta})},
\label{eq:P_fgas}
\end{multline}
where $b_\text{HSE}$ is an observational bias factor due to the assumption of hydrostatic equilibrium that was discussed in \S\ref{sec:obs_fgas}. For gas fractions measured from weak lensing plus X-ray observations the log likelihood definition is identical except that we assume the masses are unbiased, implying $b_\text{HSE}=1$ \citep[see e.g.][]{Becker2011,Bahe2012}. Note that for the likelihood of both the SMF and the cluster gas fraction we include a variance term to account for the error on the emulator prediction. This is added to avoid situations where we over-fit with respect to\ the uncertainty from the emulator alone.

The likelihood for the observational data is a combination of the likelihoods of the individual mass bins of the three data sets
\begin{multline}
            \ln\mathcal{P_{\rm likelihood}}(b_\text{cv},b_*,b_\text{HSE},\boldsymbol{\theta}) = \\
        \frac{1}{N_{\rm SMF}}\sum_{i}^{N_{\rm SMF}}\ln\mathcal{P}_{\rm SMF}(M_{*,{\rm obs},i},b_\text{cv},b_*,\boldsymbol{\theta}) + \\ 
        \frac{1}{2}\left[\frac{1}{N_{\rm HSE}}\sum_{j}^{N_{\rm HSE}} \ln\mathcal{P}_\text{gas,X-ray}(M_{500c,{\rm obs},j},b_{\rm HSE},\boldsymbol{\theta}) +\right. \\ 
        \left. \frac{1}{N_{\rm WL}}\sum_{k}^{N_{\rm WL}}\ln\mathcal{P}_\text{gas,WL}(M_{500c,{\rm obs},k}, \boldsymbol{\theta})\right], 
\end{multline}
where $N_{\rm SMF}$, $N_{\rm HSE}$ and $N_{\rm WL}$ are the number of (re-binned) observational data points (i.e.\ mass bins) for the SMF, the X-ray cluster gas fraction and the weak lensing cluster gas fraction, respectively. The values of $N$ depend on the fitted mass ranges (Table~\ref{tab:massranges}) and vary with resolution. We normalise each likelihood by the number of data points to ensure each separate likelihood is not directly dependent on the number of bins used. Furthermore, we average the likelihoods from the two types of cluster gas fraction data to ensure that the cluster gas fraction and SMF data carry equal weight. In an unweighted fit, the SMF would drive the results, because it is much better constrained. As the baryon fractions are the main driver of the baryonic suppression of the matter power spectrum \citep[see e.g.][]{VanDaalen2011,vanDaalen2020,Stijn2020,Schneider2020,Salcido2023}, we choose to give the gas fractions equal weight in our analysis.

We then combine the different likelihoods into a single posterior,
\begin{equation}
    \log \mathcal{P}_{\rm posterior} = \log\mathcal{P_{\rm likelihood}} + \log \mathcal{P}_{\rm prior},
\end{equation}
where the total prior is
\begin{align}
    \log \mathcal{P}_{\rm prior} = & \log \mathcal{P}_{\rm bias}(b_*) + \log \mathcal{P}_{\rm bias}(b_\text{cv}) + \log \mathcal{P}_{\rm bias}(b_{\rm HSE}) \nonumber \\ 
    & + \log \mathcal{P}_{\rm subgrid}(\boldsymbol{\theta}),
\end{align}
$\mathcal{P}_{\rm bias}$ are our priors for the observational bias factors, and $\mathcal{P}_{\rm subgrid}$ is our combined prior for the subgrid parameters in $\boldsymbol{\theta}$ that we wish to calibrate. For the subgrid parameters, we use flat priors that do not extend beyond the ranges used for the Latin hypercube (see Table~\ref{tab:paramtab}) in order to avoid extrapolations. The priors on the bias factors were discussed in Section~\ref{sec:obsdata}.

We also calculate the reduced $\chi^2$ for some of our models. We define the reduced $\chi^2$ as
\begin{multline}
            \chi^2_{\nu} =
            \left[\sum_{i}^{N_{\rm SMF}}\log\mathcal{P}_{\rm SMF}(M_{*,{\rm obs},i},b_\text{cv},b_*,\boldsymbol{\theta})\right. + \\ 
        \sum_{j}^{N_{\rm HSE}} \log\mathcal{P}_\text{gas,X-ray}(M_{500c,{\rm obs},j},b_{\rm HSE},\boldsymbol{\theta}) + \\ 
        \left. \sum_{k}^{N_{\rm WL}}\log\mathcal{P}_\text{gas,WL}(M_{500c,{\rm obs},k}, \boldsymbol{\theta})\right]/(N_{\rm SMF} + N_{\rm HSE} + N_{\rm WL} - N_{\theta}), 
\end{multline}

where $N_{\theta}$ is the number of sub-grid and bias parameters used for the fit.

\section{Results}\label{sec:Results}
In this section we will describe the main results from our calibration approach. We use the emulators to perform parameter sweeps in \S\ref{sec:param_sweep}, then we discuss the fitting results, first at intermediate resolution in \S\ref{sec:intresfit} and then at the other resolutions in \S\ref{sec:otherres_fit}, and finally we discuss how we use the emulator to set up two AGN feedback variations in \S\ref{sec:AGN_variations}.

\begin{figure*}
    \centering
    \includegraphics[width=\textwidth]{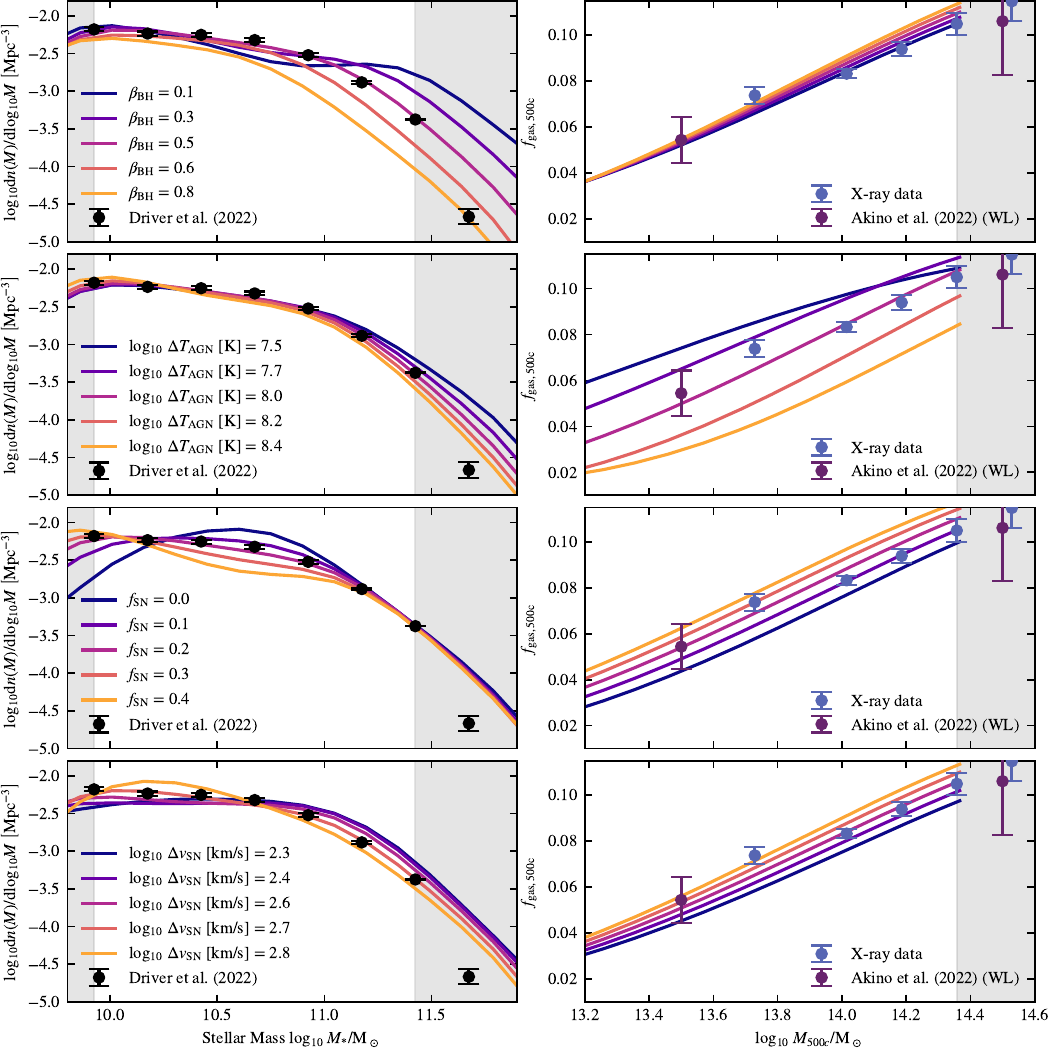}
    \caption{Subgrid parameter sweeps using the emulator trained on our 32-node Latin hypercube of $(200~\text{Mpc})^3$ intermediate-resolution simulations. The parameter sweeps are centred on the best-fitting parameters (see \S\ref{sec:intresfit}). The left and right columns show the galaxy stellar mass function and cluster gas fractions, respectively. In each row a single subgrid parameter is varied across the allowed range. From top to bottom we vary the slope of the black hole accretion rate boost factor slope, the AGN heating temperature, the stellar feedback energy, and the stellar feedback kick velocity. The grey regions indicate the mass ranges that are excluded for fitting (see also Table~\ref{tab:massranges}). Parameter sweeps help gain insight into how changes in subgrid model parameters map onto observables.}
    \label{fig:paramsweep}
\end{figure*}

\subsection{Parameter sweeps}\label{sec:param_sweep}
Emulators can be used to investigate the effect of individual parameters via parameter sweeps, where the emulator predicts the effect of varying a single parameter over the range used for the Latin hypercube, while keeping all other parameters fixed to their best-fitting values. Parameter sweeps can give valuable insight into the importance of particular physical processes and prevent calibration through emulation from becoming a black box.
The result of the subgrid parameter sweeps for our intermediate resolution runs are shown in Fig.~\ref{fig:paramsweep}. 
Looking at the response of the calibration targets, it is clear that the different parameters have distinct effects, indicating that the fits will not have any strong degeneracies between the varied subgrid parameters.

Increasing the slope of the black hole accretion rate boost factor suppresses the high-mass end of the SMF, but has almost no effect on the low-mass end and the cluster gas fractions. Increasing the AGN temperature jump leads to a mild reduction of the high-mass SMF, but a strong decrease of the cluster gas fractions. The effects of increasing the stellar feedback energy and kick velocity are more similar. In both cases the stellar masses are decreased, leading to a mass-dependent stretching of the SMF towards lower masses. Depending on the galaxy mass, the SMF can either increase or decrease, though the effect is small for the high-mass end. Cluster gas fractions decrease when either of the stellar feedback parameters increases, presumably because the stronger stellar feedback suppresses black hole growth and hence AGN feedback \citep{Bower2017}. 

\subsection{The best-fitting intermediate-resolution model}\label{sec:intresfit}
The best-fitting (i.e.\ maximum likelihood) values of the subgrid and observational bias parameters can be found in Tables \ref{tab:paramtab} and \ref{tab:biases}, respectively. These tables also list the medians and $16-84$ per cent confidence levels of the posterior distributions. 

\begin{table}
    \centering
    \caption{Results from the fitting for the observational bias factors. The second column shows the median and 16th and 84th percentiles, the third column lists the maximum likelihood value which we denote as the best-fitting.}
    \begin{tabular}{l|l|l}
        \hline
        Bias & Median+CL & best-fitting  \\
        \hline Stellar mass $\log_{10}$ $b_{\ast}$ & $0.06^{+0.11}_{-0.11}$ & 0.026 \T\B\\
        Cosmic variance $b_{\rm CV}$ & $0.98^{+0.06}_{-0.06}$ & 0.995\T\B\\
        Hydrostatic equilibrium $b_{\rm HSE}$ & $0.74^{+0.09}_{-0.09}$ & 0.743\T\B\\
        \hline
    \end{tabular}
    \label{tab:biases}
\end{table}

\begin{figure*}
    \centering
    \includegraphics[width=\textwidth]{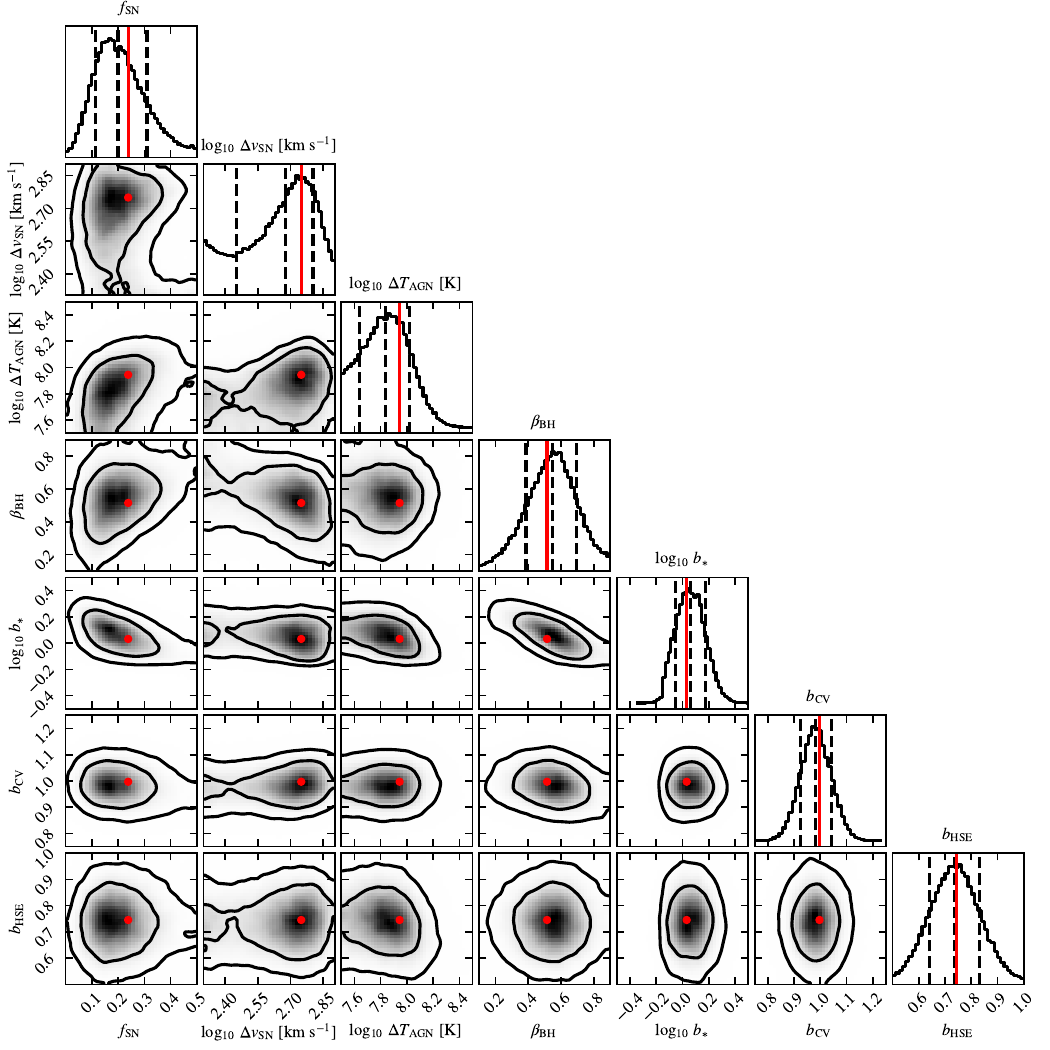}
    \caption{The posterior distributions of the model parameters resulting from fitting the emulator to the observed SMF and cluster gas fractions for intermediate-resolution simulations. The parameters shown are  the stellar feedback energy, $f_{\rm SN}$, the stellar feedback kick velocity, $\Delta v_{\rm SN}$, the AGN feedback temperature jump, $\Delta T_\text{AGN}$, the logarithmic slope of the density dependence of the black hole accretion rate boost factor, $\beta_{\text{BH}}$, the stellar mass bias, $b_{\rm M_{\rm *}}$, the hydrostatic mass bias, $b_{\rm HSE}$, and the cosmic variance bias, $b_\text{CV}$. The four subgrid parameters are described in Section~\ref{sec:simulations} and the three observational bias factors are discussed in Section~\ref{sec:obsdata}. The black contours show the $68$ and $95$ per cent confidence levels. The panels along the diagonal show the one dimensional probability density for each parameter. In these plots the three vertical lines indicate the 16th, 50th and 84th percentiles. The solid, red circles indicate the maximum likelihood values, which were used for the fiducial model. Each panel is centered on the centers of the priors given in Table~\ref{tab:paramtab}. The posteriors show that we can find a single solution that fits the simulations to the observational data.}
    \label{fig:posteriors}
\end{figure*}

The posteriors for the subgrid and bias parameters resulting from fitting the emulator predictions for intermediate-resolution simulations to the data are shown in Fig.~\ref{fig:posteriors}. The first thing to note is that the maximum likelihood model (solid, red circle) lies comfortably within the 68 per cent confidence intervals (inner contour) for each parameter and that it does not lie close to an edge of the parameter space. The chosen parameter ranges, i.e.\ the imposed priors, are thus sufficiently large for the models to bracket the target data and they do not drive the results. 

It is also clear that there are no strong degeneracies between any of the subgrid parameters or between any of the bias parameters. The absence of strongly degenerate subgrid parameters is partially by construction, because we chose to fix some of the parameters that would otherwise have caused the results to become degenerate (e.g.\ $n_\text{heat}$ and $\Delta T_{\text{AGN}}$, see \S\ref{Sec:agn_feedback}). There is, however, significant degeneracy between the slope of the density dependence of the black hole accretion boost factor ($\beta_{\text{BH}}$) and the stellar mass bias ($b_*$). These two parameters are anti-correlated. Increasing the bias shifts the observed SMF towards higher masses, which means the black hole boost factor needs to decrease to allow more stars to form in high-mass galaxies, whose growth is controlled by AGN feedback.

The best-fitting values for the galaxy mass and cosmic variance biases are $\log_{10} b_*=0.026$ and $b_\text{CV} = 0.995$, respectively. The fitted hydrostatic bias, $b_\text{HSE} = 0.743$, enables the model cluster gas fractions to agree simultaneously with the \cite{Akino2022} weak lensing data and the compilation of X-ray data. For all the bias values we find posteriors that are in agreement with the priors, so we conclude that our fitting does not put any significant additional constraints on the bias parameters.

\begin{figure*}
    \centering
    \includegraphics[width=\textwidth]{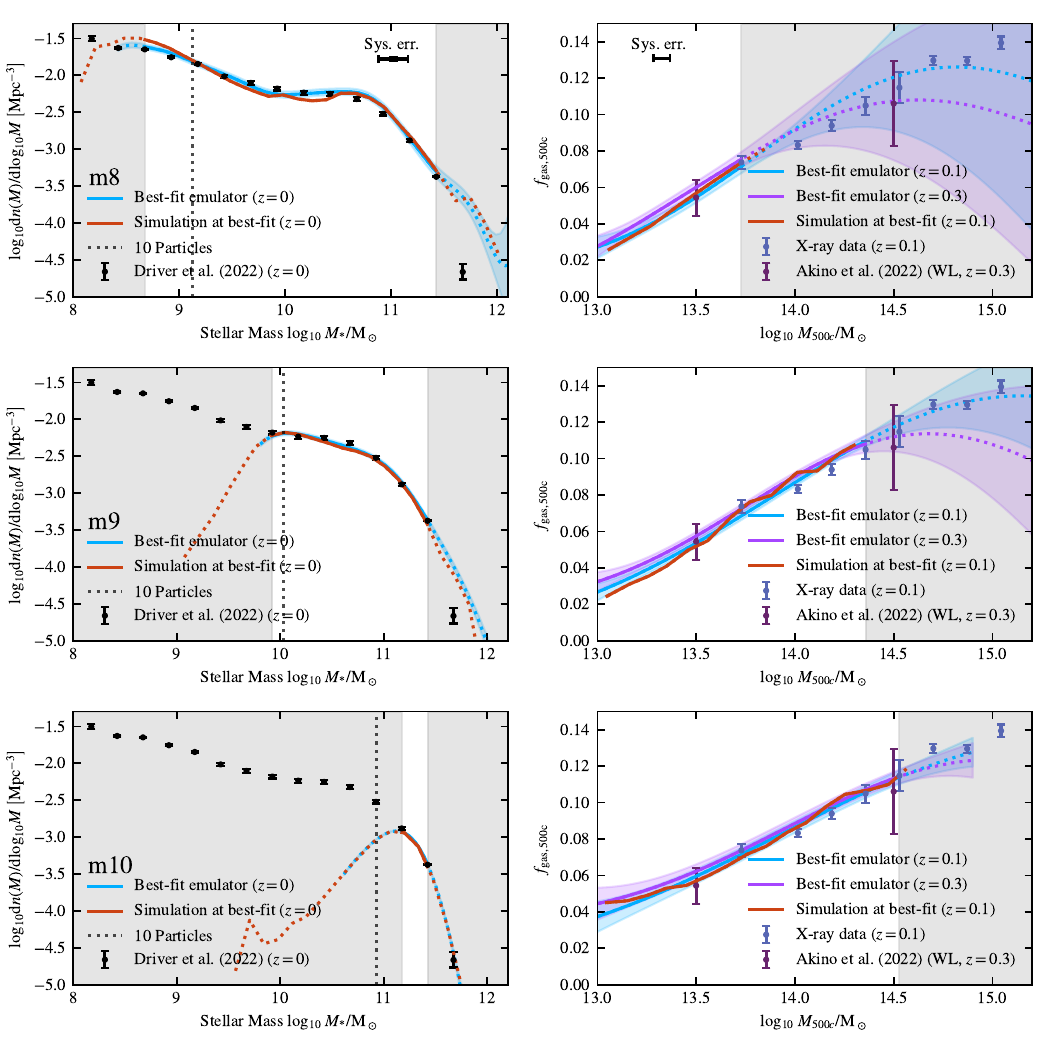}
    \caption{Comparison of the best-fitting models to the observed galaxy stellar mass function (SMF; left column) at $z=0$ and observed cluster gas fractions (right column). The top, middle and bottom rows show results for high-, intermediate- and low-resolution simulations, respectively. The observations are plotted as points with error bars (black: \citet{Driver2022} SMF at $z=0$, dark blue: compilation of X-ray data at $z=0.1$, dark magenta: \citet{Akino2022} weak lensing data at $z=0.3$). Each panel shows the best-fitting emulator prediction as a blue curve, the emulator uncertainty as a blue shaded region, and the result from a simulation using the best-fitting subgrid parameter values in a $\protect (100 \ \rm{Mpc})^3$, $(200 \ \rm{Mpc})^3$, and $(400 \ \rm{Mpc})^3$ volume for high, intermediate, and low resolution, respectively, as a red curve. For $f_{\text{gas},500c}$ we only plot the best-fitting simulation result at $z=0.1$ in red, and leave out the result at $z=0.3$ to avoid clutter. For the cluster gas fractions, besides showing in blue the $z=0.1$ emulator that should be compared with the dark blue X-ray data, we also show the $\protect z=0.3$ emulator, in magenta, that is used to fit the dark magenta \citet{Akino2022} weak lensing data. The grey regions indicate the mass ranges that are excluded from the fitting, see also Table~\ref{tab:massranges}. The model predictions are shown using dotted lines in these excluded ranges. The vertical dotted line in the left panels indicates a mass corresponding to ten stellar particles. The SMF and X-ray gas fraction data have been shifted by the best-fitting observational bias factors (see Table~\ref{tab:biases}), which are however negligible for the SMF. The SMF from the best-fitting simulation includes Eddington bias (see \S\ref{sec:eddbias}) in line with how the emulator is trained. The systematic errors given by the priors on the bias parameters are shown as points with error bars in the top panels. At each resolution we obtain excellent agreement between the emulator, a simulation with the best-fitting parameters, and the observational data.}
    \label{fig:best_fit}
\end{figure*}

The best-fitting emulator predictions for intermediate resolution are compared with the data in the middle row of Fig.~\ref{fig:best_fit}, which also shows the result of a $(200~\text{Mpc})^3$ simulation run with the best-fitting subgrid parameter values (i.e.\ our fiducial model). The left and right panels show the SMF and cluster gas fractions, respectively. The gas fractions are shown for both the redshift of the X-ray data, $z=0.1$ (light blue line and dark blue data points), and the redshift of the weak lensing data, $z=0.3$ (purple line and dark purple data points). Grey regions and dotted line styles indicate mass ranges that were excluded from the fit. The ranges can be found in Table~\ref{tab:massranges}. Note that the fitted bias factors have been used to shift the data. We obtain good agreement with the fitted observations with a reduced $\chi_\nu^2 = 1.23$ for the combined fit to the SMF and the cluster gas fractions. The good agreement between the blue and the red lines demonstrates that the emulator was able to predict accurately what the fiducial simulation would look like in the fitted mass range. 

Remarkably, the simulations fit the SMF down to galaxy masses corresponding to slightly fewer than ten stellar particles. Comparing the predicted gas fractions at $z=0.1$ and $0.3$, we see there is very little evolution. The model overshoots the gas fractions for cluster masses between $M_{500c}\approx 10^{13.8}\,\text{M}_{\odot}$ and $\approx 10^{14.5}\, \rm M_{\odot}$, by about $1\sigma$. We emphasize, however, that our observational error bars are about a factor of five smaller than the observed object-to-object scatter. Unfortunately, a box size of $(200~\rm{Mpc})^3$ (or even $(400~\rm{Mpc})^3$) is not large enough to constrain the gas fractions in haloes with $M_{500c}\geq 10^{15}\,{\rm M}_{\odot}$. Performing the same analysis in a larger volume would potentially allow the emulator to train up to the range where the $M_{500c}$-$f_\text{gas}$ relation starts to flatten.

\subsection{The best-fitting subgrid high- and low-resolution models}\label{sec:otherres_fit}
Although we use the simulation-based emulator to fit for the observational biases, the biases refer to observational effects and should thus be the same for all models. We therefore do not vary them between the different simulation resolutions. We use the intermediate-resolution simulations to fit the biases, because their resolution and box size enable us to fit a substantial mass range for both the SMF and the cluster gas fractions (see Fig.~\ref{fig:best_fit}). For the other resolutions we keep the observational biases fixed to the values listed in Table~\ref{tab:biases}. In this way we ensure that a direct comparison can be made between the three different resolutions\footnote{The \cite{Driver2022} data points at $M_\text{*,obs} \le 10^{10} \ {\rm M}_{\odot}$ were updated after we had already finished the $(2.8~\text{Gpc})^3$ intermediate-resolution FLAMINGO simulation. To be able to use the updated data for the calibration of the high-resolution simulations, which resolve the SMF down to masses for which the data were updated, we re-fit the observational biases at intermediate resolution while keeping the subgrid parameters constant. The stellar mass bias changed from $\log_{10}$ $b_* = 0.031$ to 0.026, the cosmic variance bias changed from $b_{\rm CV} = 1.014$ to 0.995 and the HSE bias from $b_{\rm HSE}=0.745$ to 0.743. The bias values changed by a negligible amount with respect to\ the 16th$-84$th percentile confidence levels, for both $b_{*}$ and $b_{\rm HSE}$ the change is less than $3$ per cent of the 16th$-84$th percentile range. For $b_{\rm CV}$ the change is $\sim15$ per cent of the 16th$-84$th percentile range. The values we report in Table~\ref{tab:biases} use the most up-to-date \cite{Driver2022} data.}.

Fixing the observational biases to the values found for intermediate resolution leaves only four parameters to fit for high resolution. For low resolution we only have two parameters to vary because we turn off stellar feedback as these simulations do not resolve the masses below which stellar feedback dominates (see \S\ref{sec:stellar_feedback}). The best-fitting parameter values for each resolution can be found in Table~\ref{tab:paramtab}. Corner plots of the posterior distributions for the subgrid parameters are shown in Appendix~\ref{sec:extra_posts}. A comparison of the best-fitting emulator prediction, the data and runs using the predicted best-fitting subgrid parameter values is shown in the top and bottom rows of Fig.~\ref{fig:best_fit} for $(100~\text{Mpc})^3$ high- and $(400~\text{Mpc})^3$ low-resolution volumes respectively. 

At high resolution there is again excellent agreement between the emulator prediction and the observed data, with reduced $\chi^2_{\nu} = 1.15$. The high-resolution simulation resolves the largest range of stellar mass in the SMF, from $\approx 10^{8.6}\, \rm M_{\odot}$ to $\approx 10^{11.5} \,\rm M_{\odot}$. There is a dip around a mass of $10^{10.2}~{\rm M}_{\odot}$ and a slight bump around the knee of the mass function. but the maximum deviation from the data is less than $5$ per cent. It seems that the emulator was unable to predict the dip, and the best-fitting simulation falls outside of the predicted errors. Comparing the predicted errors between the different resolutions, it is clear that the high-resolution simulation has the largest predicted error. This is due to it using the smallest box size. This causes the emulator prediction to be too "smooth" when compared with simulation results. The deviation at the dip is less than the $1~\sigma$ uncertainty due to cosmic variance. The small box size $(100 \ \rm{Mpc})^3$ used for calibration at high resolution, limits the mass range that can be used to fit the gas fractions to halo masses lower than $6\times 10^{13} \ \rm{M}_{\odot}$. This leaves only two data points to compare to. The agreement in the fitted range is however very good.

Comparing the best-fitting subgrid parameter values for the high-resolution model to those for intermediate resolution (Table~\ref{tab:paramtab}), we see that the stellar feedback requires about twice as much energy and about half as high a kick velocity. This reflects the need for stronger stellar feedback when higher gas densities are resolved and the fact that feedback can be efficient down to smaller wind velocities in the lower-mass haloes that remained unresolved at intermediate resolution. While the AGN heating temperatures are very similar, the high-resolution simulations require a much smaller slope of the black hole accretion rate boost factor, $\beta_{\text{BH}}= 0.038$ (where zero corresponds to no boost) versus $\beta_{\text{BH}}= 0.514$ at intermediate resolution. Since the high-resolution simulation can resolve higher gas densities, and hence higher black hole accretion rates, we do not need to boost the accretion rate as much.

At low resolution the agreement with the data is also very good, with reduced $\chi_\nu^2 = 0.95$. Now it is the stellar mass range that is very limited, $M_* \approx$ $10^{11.17} \, {\rm M}_{\odot}$ to $M_* \approx$ $10^{11.5} \, {\rm M}_{\odot}$, which includes only two data points. The larger box size of $(400~{\rm Mpc})^3$ allows for the use of the two \cite{Akino2022} weak lensing data points as well as five X-ray data points for fitting the cluster gas fractions. However, the high-mass plateau of the gas fractions remains out of reach for this box size. The comparison of the best-fitting subgrid parameter values of the low-resolution model to those of the higher-resolution simulations (Table~\ref{tab:paramtab}) is difficult to interpret because the low-resolution model requires a much lower threshold density for star formation, a much higher black hole seed mass, and does not include any stellar feedback. 

As we obtain a good fit to the same data for each of the three resolutions, we conclude that we have good 'weak convergence' between the three resolutions, using the terminology of \citet{Eaglemain}. The FLAMINGO suite includes high-, intermediate-, and low-resolution simulations that were run with our fiducial subgrid parameter values in volumes with side lengths of 1, 2.8, and 1~Gpc, respectively. For a comparison of these models with other data, we refer to Schaye et al. (2023).

\subsection{Feedback variations}\label{sec:AGN_variations}
One of the goals of FLAMINGO is to investigate the impact of feedback on cosmological observables. In this section we show how we use emulators to calibrate simulations to produce gas fractions or SMFs that have been shifted away from their fiducial, observed values. We focus mostly on changes to the gas fractions, as previous work has shown that baryon fractions in groups and clusters anti-correlate with the baryonic suppression of the matter power spectrum on the scales relevant for current and next generation surveys \citep[e.g.][]{Semboloni2013,vanDaalen2020,Stijn2020,Salcido2023}. For clusters, the gas fractions dominate over the stellar fraction when computing the baryon fractions (the stellar mass content of haloes becomes important at smaller scales). While most of our variations use our fiducial thermal AGN feedback model, we will also calibrate a model that uses kinetic, jet-like AGN feedback.

To quantify the effect of reasonable changes in the astrophysics, we include a set of  feedback variations in the simulation suite. These simulations should at least bracket the uncertainty in the cluster gas fraction data, while fitting the SMF data. Previous works created variations of subgrid physics based directly on the values of certain subgrid parameters. For example, the BAHAMAS project \citep{BAHAMAS2018} varied the AGN heating temperature by $\pm 0.2$~dex, which resulted in very small changes to the SMF and cluster gas fractions that roughly bracketed the observational uncertainty. To arrive at the values of the subgrid parameters for our runs, we make use of the emulators and we will allow all fitted subgrid parameters to vary. Our variations are based on systematically shifting of the data, based on their uncertainties, making the variations less reliant on the sub-grid model used. We also include models with gas fractions that are probably ruled out observationally, because we anticipate these will be useful to gain insight into the effect of baryonic feedback on other cosmological observables. 

The variations are run at intermediate resolution. We use the fiducial values of the observational bias factors listed in Table~\ref{tab:biases}. For the gas fraction variations, the SMF data are kept the same except for one variation, where we systematically reduce all observed stellar masses. The $f_\text{gas}$ data are shifted up  by $2\sigma$ and down by 2, 4 and 8$\sigma$ for the fgas$+2\sigma$, $-2\sigma$, $-4\sigma$ and $-8\sigma$ models respectively, where $\sigma$ is the error obtained from bootstrapping for the X-ray data, or the error on the fit for the weak lensing data from \citet{Akino2022}, as discussed in \S\ref{sec:obs_fgas}. We systematically shift all the data by $N\sigma$ under the assumption that the errors in the gas fraction are mostly systematic and correlated. We shift in steps of 2 and $4\sigma$ instead of a smaller shift (for example $1\sigma$) as the cluster-to-cluster scatter is much larger than the errors we found from bootstrapping (see \S\ref{sec:xraydata}). We also create a models that vary the SMF. As the baryonic suppression is sensitive to the total baryon fraction \citep[see e.g.][]{Salcido2023}, we include these variations to investigate the effect of changes in the baryon fraction at a constant gas fraction, and to see the effect of changing the stellar fractions. For these variations, we systematically shift the SMF data to lower masses according to the $1\sigma$ given by the stellar mass bias (0.14 dex; \S\ref{sec:mstar_sysbias}). For the M*$-1\sigma$ model we use the fiducial gas fractions and for the fgas$-4\sigma+\text{M*}-1\sigma$ model we simultaneously shift the X-ray and weak lensing gas fractions down by $4\sigma$.

The best-fitting subgrid parameter values for the feedback variations can be found in Table~\ref{tab:variations}. The changes in the subgrid parameters with respect to\ the fiducial model are small. As expected, the AGN subgrid parameters bracket the fiducial values, with the fgas$-2\sigma$ model having a slightly higher AGN feedback temperature. As could already be seen in Fig.~\ref{fig:paramsweep}, the gas fraction is very sensitive to $\Delta T_\text{AGN}$, which varies by only 0.37~dex between the fgas$+2\sigma$ and $-2\sigma$ models, in good agreement with BAHAMAS. The fgas$-4\sigma$ and $-8\sigma$ models follow this trend. Changes in the gas fractions are driven mainly by changes in $\Delta T_\text{AGN}$. Going from the fgas$-4\sigma$ to the M*$-1\sigma$ + fgas$-4\sigma$ model, the biggest change is seen in $f_{\text{SNII}}$ and $\beta_{\text{BH}}$, as expected from Fig.~\ref{fig:paramsweep}. The increase in the BH accretion boost factor is required to compensate for the removal of gas by the increased supernova energy.

\begin{figure*}
    \centering
    \includegraphics[width=\textwidth]{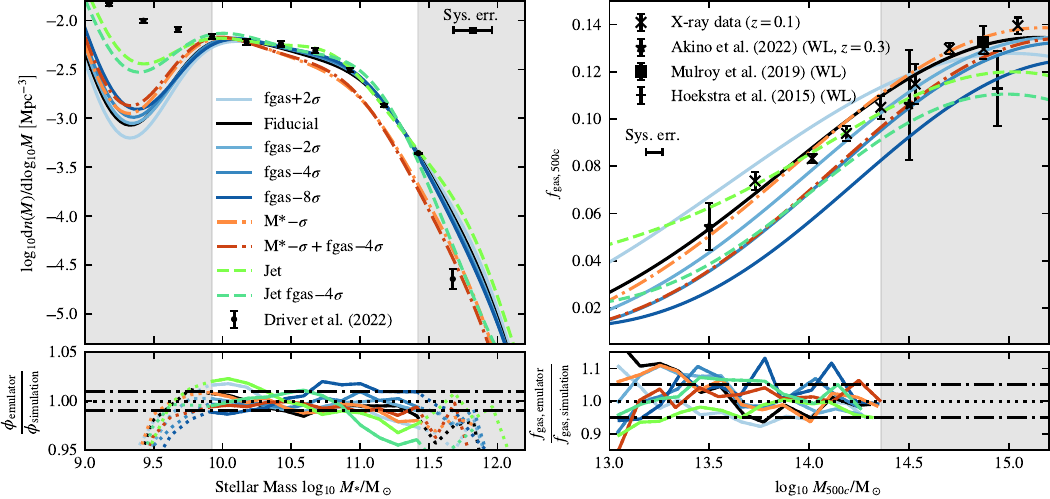}
    \caption{Top left and right panels: The emulator predictions for the SMF and gas fractions, respectively, for the feedback variations and the fiducial model (different colors, as indicated in the legend). The observations are shown as black points with error bars. In the top corners of the panels we indicate the assumed systematic errors in the data from the priors on the fitted biases. The bottom panels show the ratio of the emulator prediction and a $(200~\text{Mpc})^3$ simulation run with the same parameters. In both panels the black dotted line indicates a ratio of one. For the SMF ($f_{\text{gas},500c}$), the black dot-dashed lines indicate deviations of $1$ per cent ($5$ per cent). We only show the cluster gas fraction emulator prediction at $z=0.1$ and leave out the $z=0.3$ gas fraction results to avoid clutter. The excluded mass range for fitting is indicated by the grey regions (see also Table~\ref{tab:massranges}.) We use the emulators to make a direct mapping between our subgrid physics models and systematic shifts in the observations, based on the observational errors.}
    \label{fig:fb_vars}
\end{figure*}

The feedback models are compared with the fiducial model and the calibration data in Fig.~\ref{fig:fb_vars}. In the top two panels we show the emulator predictions for the SMF and the gas fractions for each of the variations. Within the fitted mass ranges there is excellent agreement for the SMF between all the different cluster gas fraction variations. There is good agreement between $f_\text{gas}$ for the $f_{\rm gas}-4\sigma$ and the SMF$-1\sigma$ + $f_{\rm gas}-4\sigma$ variations. In the bottom panels we compare the emulator predictions to the results of $(200~\text{Mpc})^3$ simulations run with the best-fitting parameters. For the SMF, we see that the emulator predictions are accurate at around the per cent level, with only the jet model fgas$-4\sigma$ deviating by $\approx5$ per cent. For $f_\text{gas}$, all predictions are accurate to $\approx 10$ per cent, and most predictions are accurate to within $\approx 5$ per cent. The accuracy is slightly better than the expected emulator accuracy from cross-checks (see Table~\ref{tab:acc_table}).
We conclude that by allowing for small adjustments to four subgrid parameters, we are able to vary specific observables while keeping others constant.

\begin{table*}
    \centering
    \caption{best-fitting values for the subgrid parameters for the feedback variations at intermediate resolution. The columns list 
    the name of the variation, the number of $\sigma$ by which the observed $f_\text{gas}$ data was shifted, and for each parameter the median and 16th to 84th percentile confidence level (CL), and the best-fitting (i.e.\ maximum likelihood) fiducial values. Note that for the jet AGN model the seventh and eighth columns show $v_{\text{jet}}$ instead of the heating temperature, while for the other feedback variations they show $\Delta T_{\rm AGN}$.}
    \begin{tabular}{l|r|l|l|l|l|l|l|l|l|l}
        \hline
         &  & \multicolumn{2}{c}{$f_{\rm SN}$} & \multicolumn{2}{c}{$\Delta v_{\rm SN}$ [km~s$^{-1}$]} & \multicolumn{2}{c}{$\Delta T_{\rm AGN}$ [K] or $v_{\text{jet}}$ [km~s$^{-1}$]} & \multicolumn{2}{c}{$\beta_{\text{BH}}$}  \\
        Variation & $\sigma$ & Median+CL & best-fitting & Median+CL & best-fitting & Median+CL & best-fitting & Median+CL & best-fitting \\
        \hline
        fgas$+2\sigma$ & +2 & $0.22^{+0.09}_{-0.08}$ & $0.219$ & $525^{+151}_{-186}$ & 577 & $10^{7.69^{+0.16}_{-0.13}}$ & $10^{7.71}$ & $0.58^{+0.10}_{-0.10}$ & $0.554$ \T\B\\
        Fiducial & 0 & $0.20^{+0.11}_{-0.09}$ & $0.238$ & $479^{+167}_{-197}$ & 562 & $10^{7.84^{+0.18}_{-0.20}}$ & $10^{7.95}$ & $0.55^{+0.15}_{-0.16}$ & $0.514$\\
        fgas$-2\sigma$ & -2 & $0.21^{+0.08}_{-0.07}$ & $0.206$ & $478^{+149}_{-179}$ & 552 & $10^{8.03^{+0.14}_{-0.16}}$ & $10^{8.08}$ & $0.54^{+0.10}_{-0.09}$ & $0.497$\T\B\\
        fgas$-4\sigma$  & -4 & $0.20^{+0.08}_{-0.07}$ & $0.191$ & $479^{+167}_{-162}$ & 532 & $10^{8.18^{+0.13}_{-0.13}}$ & $10^{8.21}$ & $0.51^{+0.09}_{-0.09}$ & $0.482$\T\B\\
        fgas$-8\sigma$  & -8 & $0.15^{+0.07}_{-0.06}$ & $0.145$ & $417^{+156}_{-154}$ & 483 & $10^{8.36^{+0.09}_{-0.11}}$ & $10^{8.40}$ & $0.49^{+0.07}_{-0.08}$ & $0.462$\T\B\\
        M*$-\sigma$  & 0 & $0.30^{+0.10}_{-0.10}$ & $0.322$ & $537^{+124}_{-198}$ & 608 & $10^{7.98^{+0.14}_{-0.17}}$ & $10^{8.06}$ & $0.68^{+0.11}_{-0.10}$ & $0.626$\T\B\\
        M*$-\sigma$ + fgas$-4\sigma$  & -4 & $0.25^{+0.10}_{-0.08}$ & $0.261$ & $490^{+127}_{-174}$ & 557 & $10^{8.25^{+0.13}_{-0.13}}$ & $10^{8.27}$ & $0.65^{+0.09}_{-0.09}$ & $0.620$\T\B\\
        Jet & 0 & $0.19^{+0.07}_{-0.06}$ & $0.166$ & $562^{+196}_{-164}$ & 477 & $977^{+311}_{-236}$ & 836 & $0.54^{+0.10}_{-0.12}$ & $0.597$\T\B\\
        Jet + fgas$-4\sigma$ & -4 & $0.18^{+0.08}_{-0.06}$ & $0.176$ & $524^{+200}_{-162}$ & 527 & $1949^{+238}_{-251}$ & 1995 & $0.44^{+0.07}_{-0.08}$ & $0.439$
        \T\B\\\hline
    \end{tabular}
    \label{tab:variations}
\end{table*}

In addition to the parameter variations, we also calibrate a different implementation of AGN feedback. As described in \S\ref{sec:jet_feedback} this model uses kinetic bipolar kicks instead of thermal injections to distribute AGN feedback energy around accreting BHs\footnote{Due to a bug, the calibration of the jet models was done using a version of
the model where the jets are launched along the z-direction of the simulation
box, instead of along the spin axis of the black hole. We have verified that this
leads to small differences, in agreement with the results reported by \citet{Husko2023jetdir}, who showed that the directionality of the jets has little effect.
When using the correct implementation, the agreement with the emulator of
the SMF becomes slightly better for both runs that use jets; and for fgas, the
agreement only worsens outside the range used for calibration.}. As the subgrid model differs fundamentally from the fiducial model, we run a new Latin hypercube with 32 intermediate-resolution simulations in $(200~\text{Mpc})^3$ volumes. The subgrid parameter ranges for this hypercube can be found in Table~\ref{tab:jet_params}. To construct the emulator, we again follow the prescription of Section~\ref{sec:emulator_descr} and we again verify its accuracy using cross-checks (see Table~\ref{tab:acc_table}). The goal is to have a simulation with a different implementation of AGN feedback calibrated to the same observables as the fiducial implementation. We therefore use the same fitting limits, methods and likelihoods as for the fiducial intermediate-resolution model. For the jet model we fit to both the fiducial data and to the perturbed data used to calibrate the $f_{\rm gas}-4\sigma$ model. The resulting medians and best-fitting values can be found in Table~\ref{tab:variations}.

The jet models are shown as the green lines in Fig.~\ref{fig:fb_vars}. They show some differences from the fiducial thermal AGN feedback models. The jet models fit the knee of the SMF slightly better by having slightly more galaxies with $M_{*}\approx10^{10.7} \,\text{M}_{\odot}$. The difference at the very low-mass end of the SMF, below the fitted range, is due to the fact that the bug in the threshold of star formation for zero metallicity gas (see footnote \ref{footnote:sf_bug}) was fixed for the jet models. The $f_{\text{gas}}-4\sigma$ jet model also has a significant reduction in the number of galaxies with masses above our fitting limit, thus yielding a SMF with a steeper high-mass cut off. However, the bottom panel suggests that this is at least partially explained by the fact that the emulator under-predicts the number density by a few per cent. Compared with the thermal AGN models fit to the same data, the jet models predict higher gas fractions in groups ($\text{M}_{500c}\sim10^{13}\,\text{M}_{\odot}$), where there is, however, no observational data. From the bottom panels we can see that for $f_\text{gas}$ the accuracy of the jet emulator does not differ significantly from the emulator for the thermal AGN feedback models.

\section{Conclusions}\label{sec:conc}
In order to fully exploit the large-scale structure data that will become available with surveys like \textit{Euclid} and LSST, we need to acquire a deeper understanding of how baryonic effects, like AGN and stellar feedback, impact the matter distribution. The most self-consistent way of experimenting with these effects is through the use of cosmological hydrodynamical simulations. The FLAMINGO project provides such simulations in volumes sufficiently large to study the evolution of large-scale structure and massive galaxy clusters for different numerical resolutions, cosmologies and astrophysical models.

As feedback processes originate on unresolved scales, we have to add them via subgrid prescriptions. However, because these subgrid models are theoretically not well constrained, they need to be calibrated to reproduce a relevant set of observables. Previous simulation projects like EAGLE \citep{Eaglemain,EagleCal}, IllustrisTNG \citep{Illustris2018}, BAHAMAS \cite{BAHAMAS2017,BAHAMAS2018} and SIMBA \cite{SIMBA2019} achieved good agreement with data by varying subgrid parameters by hand until the simulation lined up with the target observations. However, for cosmology a more robust and objective calibration method is desirable, particularly if it can also be used to predict the effect of subgrid variations that have not been simulated directly.

To create a robust method of calibration, we make use of machine learning, specifically Gaussian process emulators. Instead of emulating the effects of changes in the cosmological parameters, which is becoming a common application of machine learning in cosmology, we emulate the observables that we want to match to observations as a function of a set of subgrid parameters. For three different numerical resolutions, which span a factor of 64 in particle mass, we train an emulator on 32 input simulations where we vary the four most impactful subgrid parameters, two of which relate to stellar feedback and two of which relate to AGN feedback (Section~\ref{sec:simulations}). In addition, we train an emulator for another intermediate-resolution implementation of AGN feedback, which uses jets (i.e., directed kinetic feedback) instead of injecting the feedback energy thermally. At each resolution we run simulations with $360^3$ gas particles, implying a $(100~\text{Mpc})^3$, $(200~\text{Mpc})^3$ and $(400~\text{Mpc})^3$ volume for FLAMINGO high [m8], intermediate [m9] and low [m10] resolution, respectively. We then use MCMC to fit the emulator to carefully selected observational data. We repeat the same procedure for each resolution, and only change the fitted mass ranges to account for resolution and box size limitations. Additionally, we have created a set of subgrid physics implementations based on fitting the emulators to the data after systematically shifting it by $N\sigma$. 

We calibrate to the observed low-redshift galaxy stellar mass function (SMF) from the GAMA survey and a compilation of group and cluster gas fraction measurements based on X-ray and weak lensing data. A novel aspect of our approach is that we also fit for possible observational biases (i.e., systematic errors). We account for biases in the stellar mass and the cluster mass inferred from X-ray data under the assumption of hydrostatic equilibrium, as well as for the effect of cosmic variance on the SMF. In addition, we account for the effect of random errors in the observed stellar mass on the SMF (i.e., Eddington bias) by randomly perturbing the simulated stellar masses(Section~\ref{sec:obsdata}). The observational biases are only fit during the calibration of the intermediate-resolution simulations and the best-fitting values are then also applied to the other resolutions. 

Our main conclusions are:
\begin{enumerate}
    \item By carefully setting up the subgrid parameter space, we were able to train emulators that are more accurate than the target observational constraints (Fig.~\ref{fig:hyper_perform}).
    \item The emulator framework enables simultaneously fitting for subgrid parameters and observational biases. For FLAMINGO, the posteriors found for the biases are driven by and in agreement with the priors. We find a negligible value for the stellar mass and cosmic variance error, and a hydrostatic bias of $b_{\rm HSE} = 0.743$.
    \item Emulators can be used to make parameter sweeps, i.e.\ plots showing how the trained relation depends on the value of a single subgrid parameter (Fig.~\ref{fig:paramsweep}). As the emulators give the continuous response of the trained relation to changes in subgrid parameters, emulators can be used to gain a deeper understanding of how the observable relations are affected by the subgrid models.
    \item The parameter space that we explore is devoid of major degeneracies between the subgrid parameters. The emulator+MCMC framework finds a single best-fitting solution (Fig.~\ref{fig:posteriors}). We note that this is partially by construction, as parameters that had major degeneracies were omitted from the parameter space (see Section \ref{sec:simulations}). For future work it might be interesting to see if these degeneracies can be solved by fitting the model to additional observational data.
    \item At each resolution we find excellent agreement between the best-fitting model and the calibration data (Fig.~\ref{fig:best_fit}). 
    \item The emulator framework can be used to map observational uncertainties onto changes in subgrid parameters. By fitting the emulator to variations in gas fractions and the SMF, we produce a set of simulations for which specific observables are varied while keeping others constant (Fig.~\ref{fig:fb_vars}). As the model variations are directly tied to observations, the resulting simulations can be used to quantify the effect of uncertainties in the calibration data on the predictions for other observables. 
    \item We used the emulator framework to calibrate a different implementation of the model, which we did for kinetic AGN feedback (in contrast with the thermal AGN feedback used our fiducial model; Fig.~\ref{fig:fb_vars}). By making different models match the same calibration observations, the simulations can be used to quantify the uncertainty in predictions for other observables due to uncertainties in the underlying physics.
\end{enumerate}

We have used Gaussian process emulators to create a close link between subgrid models and observations. By creating a robust statistical framework for calibration, future hydrodynamical simulations will be able to use available and upcoming data to constrain the subgrid physics and to quantify the uncertainty in the predictions of simulations that remains after the models have been constrained to fit particular sets of data. In this work we have focused on calibrating simulations using different resolutions, and a single variation of the implementation of AGN feedback. For future work the same framework could be used to get agreement between different simulation codes and subgrid models for specific observables. In this way we could improve our understanding of the degeneracies between different methods and the uncertainties in their predictions.

In the companion paper \citet{Flamingomain} we present the large-volume FLAMINGO simulations that use the calibrated parameter values that we obtained here. More information on and visualisations of the FLAMINGO simulations can be found on the website.\footnote{\url{https://flamingo.strw.leidenuniv.nl/}}

\section*{Acknowledgements}
This work is partly funded by Vici grant 639.043.409, Veni grant 639.041.751 and research programme Athena 184.034.002 from the Dutch Research Council (NWO). This work used the DiRAC@Durham facility managed by the Institute for Computational Cosmology on behalf of the STFC DiRAC HPC Facility (www.dirac.ac.uk). The equipment was funded by BEIS capital funding via STFC capital grants ST/K00042X/1, ST/P002293/1, ST/R002371/1 and ST/S002502/1, Durham University and STFC operations grant ST/R000832/1. DiRAC is part of the National e-Infrastructure.EC is supported by the funding from the European Union’s Horizon 2020 research and innovation programme under the Marie Skłodowska-Curie grant agreement No 860744 (BiD4BESt). We gratefully acknowledge financial support from the Swiss National Science Foundation (SNSF) under funding reference 200021\_213076. F. H. would like to acknowledge support from the Science Technology Facilities Council through a CDT studentship (ST/P006744/1). I.V. gratefully acknowledges UKRI (EP/W011956/1) and Wellcome (218261/Z/19/Z) funding. ARJ, CGL, JCH and CSF acknowledge the STFC consolidated grants ST/T000244/1
and ST/X001075/1. This research project has received funding from the European Research Council (ERC) under the European Union's Horizon 2020 research and innovation programme (grant agreement No 769130). The research in this paper made use of the SWIFT open-source simulation code
(\url{http://www.swiftsim.com}, \cite{SWIFT_ascl}) version 0.9.0.

\section*{Data Availability}

The \textsc{swift-emulator} framework used for this work is publicly available, see \citep{Kugel2022}\footnote{\url{https://swiftemulator.readthedocs.io/en/latest/}}. The simulation data used will be provided upon reasonable request to the corresponding author.

\bibliographystyle{mnras}
\bibliography{example} 




\appendix

\section{Different apertures} \label{sec:ap_aper}
\begin{figure}
    \centering
    \includegraphics[width=0.5\textwidth]{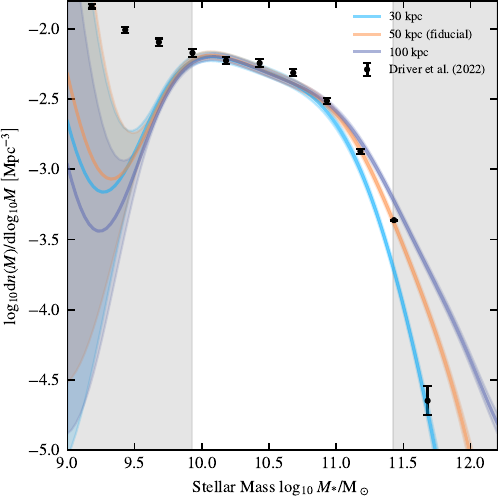}
    \caption{The effect on the SMF of choosing a different aperture when measuring stellar masses in the simulation. For each line we set up a new emulator based on the simulation results for the corresponding aperture. Each emulator is then used to predict the behaviour at the best-fitting parameter values for the fiducial 50~kpc aperture. Differences between the apertures start to occur above a stellar mass of $10^{11}~\rm{M}_{\odot}$}.
    \label{fig:apertures}
\end{figure}

Fig.~\ref{fig:apertures} compares the SMF results for different choices of 3D apertures with radii of 30, 50 (our fiducial aperture) and 100~kpc. For each non-fiducial aperture we retrain the emulator on the SMFs obtained with the different aperture. The new emulator, based on a different aperture, is then evaluated at the fiducial subgrid parameter values. We do not refit the SMF for each aperture, because we wish to quantify the effect of the aperture size on the SMF predicted by a given simulation. The choice of aperture only has an impact at the largest stellar masses (see also \citealt{Eaglemain}). For our analysis this implies that the main effect of an increase in aperture would be a slight increase of the slope of the density dependence of the AGN accretion rate boost factor. However, for the fitted mass range this effect is relatively small. The effect of using a mass measurement method more similar to that used by observers may be larger \citep[e.g.][]{deGraaff2022}, but such a comparison is not feasible at the resolution of our simulations.

\section{Posteriors for high- and low-resolution}\label{sec:extra_posts}
\begin{figure}
    \centering
    \includegraphics[width=0.5\textwidth]{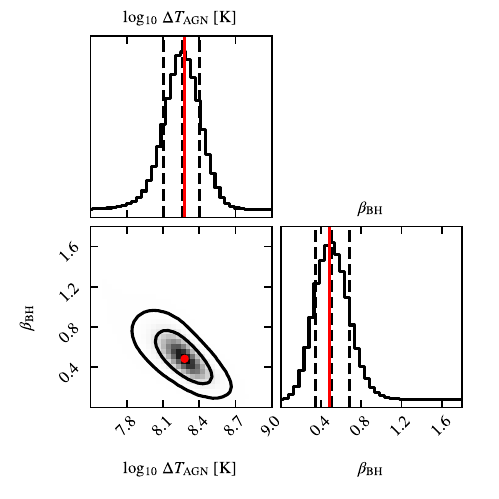}
    \caption{The posterior distributions of the model parameters resulting from fitting the emulator for low-resolution simulations to the observed SMF and cluster gas fractions. The parameters shown are the AGN feedback temperature jump $\Delta T_\text{AGN}$ and the logarithmic slope of the density dependence of the black hole accretion rate boost factor, $\beta_{\text{BH}}$. The two subgrid parameters are described in Section~\ref{sec:simulations}. The black contours show the $68$ and $95$ per cent confidence levels. The panels along the diagonal show the one dimensional probability density for each parameter. In these plots the three vertical lines indicate the 16th, 50th and 84th percentiles. The solid, red circle indicate the maximum likelihood values, which were used for the fiducial model. There is some degeneracy, but there is a clear single best-fitting solution.}
    \label{fig:posteriorslow}
\end{figure}

The posteriors for low resolution are shown in Fig.~\ref{fig:posteriorslow}. There is a degeneracy between the two parameters. Both parameters are sampled well within our chosen ranges. Even though the range for the heating temperature is much wider than for the other resolutions, we find that the best-fitting value is in the range where AGN feedback is well sampled, and does not suffer from catastrophic numerical overcooling (see §\ref{Sec:agn_feedback}). 

\begin{figure*}
    \centering
    \includegraphics[width=\textwidth]{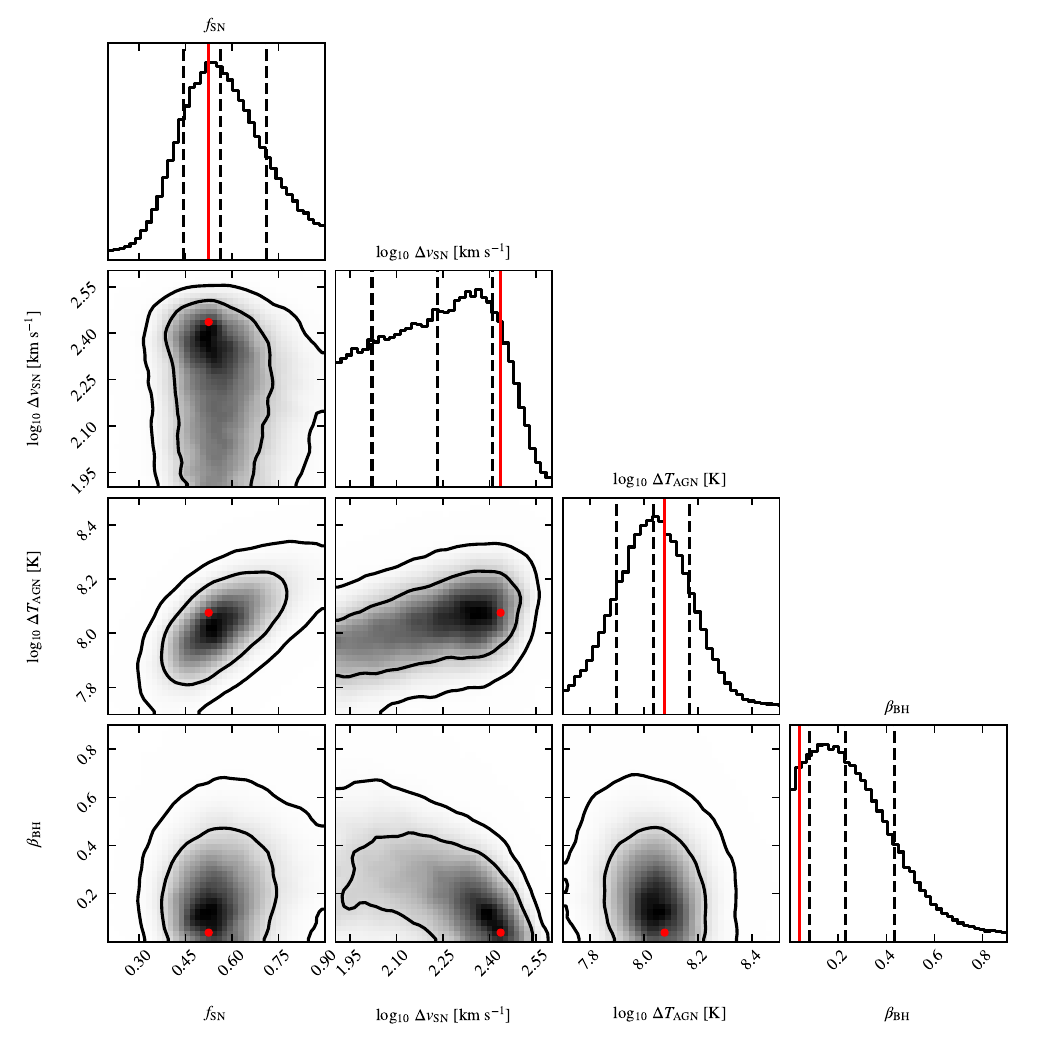}
    \caption{The posterior distributions of the model parameters resulting from fitting the emulator for high-resolution simulations to the observed SMF and cluster gas fractions. The parameters shown are  the stellar feedback energy, $f_{\rm SN}$, the stellar feedback kick velocity, $\Delta v_{\rm SN}$, the AGN feedback temperature jump, $\Delta T_\text{AGN}$ and the logarithmic slope of the density dependence of the black hole accretion rate boost factor, $\beta_{\text{BH}}$. The four subgrid parameters are described in Section~\ref{sec:simulations}. The black contours show the $68$ and $95$ per cent confidence levels. The panels along the diagonal show the one dimensional probability density for each parameter. In these plots the three vertical lines indicate the 16th, 50th and 84th percentiles. The solid, red circles indicate the maximum likelihood values, which were used for the fiducial model. The results show some moderate degeneracies, but the individual parameters each have a clear peak close to the best-fitting values.}
    \label{fig:posteriorshigh}
\end{figure*}

The posteriors for the high-resolution simulation are shown in Fig.~\ref{fig:posteriorshigh}. Similar to the intermediate-resolution posteriors we find a best-fitting model within the chosen parameter ranges. The best-fitting value for $\beta_{\rm BH}$ is quite close to the edge, partly due to a degeneracy between $\beta_{\rm BH}$ and $\Delta v_{\rm SN}$. The high-resolution posteriors are more degenerate than for intermediate-resolution. This is likely due to the fact that we fit a much broader range of the SMF, making it more important to get the balance between stellar and AGN feedback right. The posteriors show that there are some significant degeneracies in how this problem can be solved. Note that for both and high  and low resolution we have fixed the biases to the values for for intermediate resolution, see \S\ref{sec:intresfit}.

\section{Parameter ranges for the AGN Jet model}
The subgrid parameter ranges for the Latin hypercube that was used to train the emulators for the AGN jet model can be found in Table~\ref{tab:jet_params}. 
\begin{table}
    \centering
     \caption{Subgrid parameter ranges for the Latin hypercube used to train the jet model emulators.}
    \begin{tabular}{l|l}
    \hline
        Parameter & Prior \\ \hline
        $f_{\rm SN}$ & [0.0,0.5]\\
        $\Delta v_{\rm SN}$ [km~s$^{-1}$] & [$10^{2.3}$,$10^3$]\\
        $v_{\text{jet}}$ [km~s$^{-1}$] & [$10^{2.7},10^{3.5}$]\\
        $\beta_{\text{BH}}$ & [0.1,0.7]\\
        \hline
    \end{tabular}
 
    \label{tab:jet_params}
\end{table}

\bsp	
\label{lastpage}
\end{document}